\documentclass[a4paper,fleqn]{cas-dc}

\usepackage{amsfonts}
\usepackage{graphicx}
\usepackage{amsmath}
\usepackage[T2A,T1]{fontenc}
\usepackage[utf8]{inputenc}
\usepackage[ukrainian, icelandic,english]{babel}
\usepackage{braket}
\usepackage{esint}
\usepackage{hyperref}
\usepackage{mathtools}
\usepackage{amssymb}
\usepackage{systeme}
\usepackage{setspace}
\usepackage{upgreek}
\usepackage{xcolor}
\usepackage{graphicx}
\usepackage{bm}
\usepackage[sort&compress, numbers]{natbib}

\begin{document}
\shorttitle{Influence of the particle distribution}
\shortauthors{D.~Yaremchuk et al.}

\title [mode = title]{Magnetostriction in the magneto-sensitive elastomers with inhomogeneously magnetized particles: pairwise interaction approximation}

\author[1]{D.~Yaremchuk}[orcid= 0000-0003-2888-5878]
\cormark[1]
\ead{yaremchuk@icmp.lviv.ua}

\address[1]{Institute for Condensed Matter Physics of the National Academy of Sciences of Ukraine, 1, Svientsitskii Str., 79011, Lviv, Ukraine}

\author[2,3]{D.~Ivaneyko}

\address[2]{Leibniz-Institut für Polymerforschung Dresden e. V., Hohe Straße 6, 01069 Dresden, Germany}

\author[1]{J.~Ilnytskyi}

\address[3]{Institute of Applied Physics, Technische Universität Dresden, 01062 Dresden, Germany}

\cortext[cor1]{Corresponding author}

\begin{abstract}
We analyze the magnetostriction effect occurring in the magneto-sensitive elastomers (MSEs) containing inhomogeneously magnetized particles. As it was shown before, the expression for the interaction potential between two magnetic spheres, that accounts for their mutual inhomogeneous magnetization, can be obtained from the Laplace equation. We use this potential in the approximation formula form to construct magnetic energy of the sample in terms of the pairwise interactions of the particles. We show that this form of magnetic energy leads to the same demagnetizing factor as predicted by the continuum mechanics, confirming that only dipole-dipole magnetic interactions are important on a large scale. As the next step, we examine the role played by the particles arrangement on the magnetostriction effect. We consider different spatial distributions of the magnetic particles: a uniform one, as well as several lattice-type distributions (SC, BCC, HCP and FCC arrangements). We show that the particles arrangement affects significantly the magnetostriction effect if the separation between them became comparable with the particles' dimensions. We also show that, typically, this contribution to the magnetostriction effect is of the opposite sign to the one related with the initial elastomer shape. Finally, we calculate the magnetostriction effect using the same interaction potential but expressed in a form of a series expansion, qualitatively confirming the above findings.
\end{abstract}

%\begin{graphicalabstract}
%\includegraphics{figs/grabs.pdf}
%\end{graphicalabstract}

%\begin{highlights}
%\item Research highlights item 1
%\item Research highlights item 2
%\item Research highlights item 3
%\end{highlights}


\maketitle

\section{Introduction}\label{I}

The phenomenon when the magneto-responsive material changes its shape upon application of an external magnetic field, is termed ``the magnetostriction effect'' \cite{Joule47, Ginder2000,Ivaneyko14, Biller2014a, Dirk}. Studies of this phenomenon acquired new momentum in advent of the interest in composite materials \cite{Rigbi83, Jolly96, Carlson2000}, started a few decades ago. A class of magneto-sensitive elastomers (MSEs) includes composite materials containing the elastic and magnetic subsystems and, therefore, can be reversibly controlled via an external magnetic field. It turns out that uniaxial magnetostriction for MSEs is three orders of magnitude larger than in pure metals and comparable or higher than ``giant magnetostriction'' in rare-earth alloys~\cite{Bednarek2006,Koon71,Clark72}. This and others unusual mechanical properties of MSEs made them candidates for application in different devices, e.g., sensors, dampers and actuators~\cite{Koon91, Raju2018, Li2014, Kang2020}. In that practical context, prediction of the mechanical properties for these novel materials based on the details of their physical and structural properties  is of a great importance.

The question of: how exactly the inner structure of MSE influences its macroscopic behavior is a challenging theoretical task, which requires thorough description of its both elastic and magnetic subsystems. The properties of the elastic component viewed from a microscopic perspective are discussed in Ref.~\cite{Flory1985, Erman1997, Urayama2006} and literature therein. More recently, novel technics, e.g, graph theory~\cite{Amamoto2020} and machine learning~\cite{Jackson2019} were used to study their structure-property relationship.
On the other hand, the effects of the structural order in the magnetic subsystem on the macro-scale response of the material should also be studied. Relevantly, the investigations of sample properties affected by the magnetic filler distribution~\cite{Ivaneyko14, Ivaneyko2011, Ivaneyko2012, Dirk19, Dirk2022} are performed in recent years.
The most common choice for the form of the magnetic energy is the dipole-dipole interaction~\cite{Ivaneyko14, Dirk19, Jolly96, Zhang2008}. However, it is being argued in the literature~\cite{Biller2014a, Biller2014} that at the inter- particle separations comparable with their dimensions, the inhomogeneous bulk magnetization effects become important. Additionally, it was shown~\cite{Yaremchuk2020,Biller2014}, that inhomogeneously magnetized particles models and the microscopic approach based in dipole-dipole interaction predict different angular dependencies for the forces between pair of particles. As a result, one may expect somehow different predictions of how exactly the magnetic particles distribution inside an elastomer matrix influence its macroscopic response on the magnetic field.

Yet another aspect, as was shown in a similar problem of inhomogeneously polarizable spheres~\cite{Bossis93}, is a non-additivity of the interaction potential and importance of the three- particle interactions. As a result, the dipole-dipole interaction model is applicable to the MSEs in the limit of essentially diluted magnetic filler. At the same time, a general theory which takes into account the inhomogeneous magnetization of many particles may be impractically complex.

In the current paper we generalize previous magnetostriction studies of different spatial distributions of magnetic particles inside an MSE~\cite{Dirk19} towards the case of the inhomogeneously magnetizable spheres. The equilibrium strain of the spheroidal sample in the external magnetic field is considered as a measure of magnetostriction effect. Dimensionless coefficient (magnetostriction factor), that defines the sign of the strain, can be split into two terms: one is related to the initial shape of the elastomer, whereas the other one accounts for the initial distribution of the magnetic particles~\cite{Dirk19, Ivaneyko14}. The magnetostriction factor is evaluated for our model by using the expression for a pairwise potential energy obtained earlier~\cite{Jeff, Biller2014a, Biller2014} by solving the Laplace equation. In our study, we use two forms of this potential energy: the explicit expression for the first $n=9$ terms of the series expansion~\cite{Jeff}, and the numerical approximation based on the first $n=100$ terms~\cite{Biller2014}. The obtained results are compared with the predictions of the dipole-dipole interaction model~\cite{Dirk19} and in each case the magnetostriction factor is calculated assuming the affine deformation of an MSE sample under external magnetic field. This comparison is aimed on finding the range of applicability for the dipole-dipole interaction model for the magnetostriction effect. We are also interested in the question of which results, provided by the dipole-dipole model, hold after the inhomogeneous magnetization effect is taken into account in the pairwise interaction approximation.

Outline of the paper is as follows.
In the Sec.~\ref{II} we derive the magnetic energy of the elastomer within the pairwise interaction approximation (PIA).
In the Sec.~\ref{III} we show, that for macroscopic samples the shape dependent magnetostriction is described well by the dipole interactions, while other terms in the potential energy can be neglected.
In the Sec.~\ref{IV} the distribution dependent magnetostriction effect is considered for the uniform and the lattice-like arrangement of the particles. The effects of inhomogeneous magnetization are analyzed by comparing the magnetostriction factors with dipole-dipole model prediction.
In the Sec.~\ref{V} we analyze briefly the equilibrium strain for various values of external magnetic field and sample parameters such as its initial shape and the volume fraction of magnetic filler. The study is eventuated by Conclusions.

\section{Magnetic energy density and equilibrium strain}\label{II}

Magnetic energy of two equally sized linearly magnetizable spheres can be obtained by solving the Laplace equation~\cite{Jeff, Biller2014a, Biller2014}. The solution has a form of infinite series~\cite{Yaremchuk2020, Biller2014a}
\begin{equation}\label{LES}
U^{\mathrm{int}}_{\mathrm{LE}}=2U_0\sum_{n=3}^{\infty}\bigg(\frac{a}{r}\bigg)^n(a_n\cos^2{\theta}+b_n),
\end{equation}
where $a$ is the radius of particles and $a/r$ is dimensionless inverse distance between their centers. The angle between the direction of the external magnetic field $\bm{H}_0$ and the radius-vector $\bm{r}$ is denoted by $\theta$. The coefficients $a_n$, $b_n$ are constants, they are calculated analytically in Ref.~\cite{Jeff} up to $n=9$. The first four of them are: $a_3=3\beta_1$, $b_3=-\beta_1$, $a_{4,5}=b_{4,5}=0$, $a_6=3\beta_1^2$, $b_6=\beta_1^2$.

These series is found to converge rather slowly and, therefore, it was suggested to keep the terms up to $n=100$~\cite{Biller2014a, Biller2014} and to approximate it with the following expression
\begin{equation}\label{Eq1}
U^{\mathrm{int}}=2U_0\sum_{k=3}^{7}\! \beta_1^{p_k-1} \bigg[     \frac{a^kA_k}{(r-aB_k)^k}+\frac{a^kC_k}{(r-aD_k)^k}\cos^2{\theta}   \bigg],
\end{equation}
where the exact values of the coefficients $A_k$, $B_k$, $C_k$ and $D_k$ are given in the Table~\ref{tab:table1} (from~\cite{Biller2014}).

The coefficient $U_0$ in Eqs.~(\ref{LES},~\ref{Eq1}) is proportional to the magnetic energy of an isolated sphere under external field
\begin{equation}\label{Eq2}
U_0=-\frac{3}{2}\mu\mu_0\beta_1H^2_0V_p,
\end{equation}
where $\mu=\mu_{\mathrm{medium}}/\mu_0$ is the relative permeability of a medium and $\mu_0$ is the permeability of a vacuum. In the case of MSE, the medium is typically non-magnetic~\cite{Jolly96, Carlson2000} therefore, one can set $\mu=1$. The factor $\beta_n=\frac{n\chi}{n\chi+2n+1}$ is the function of magnetic susceptibility $\chi$. The factor $V_p$ is the volume of a particle $V_p=\frac{4\pi}{3}a^3$.
\begin{table}
\centering
\caption{Coefficients entering the expression for the potential energy~\cite{Biller2014}}
\label{tab:table1}
\begin{tabular}{ c|c|c|c|c|c }
\hline\hline
 k&  $A_k$ & $B_k$  &   $C_k$                         &        $D_k$       &     $p_k$ \\
\hline\hline
 3& -1     & 0      &        $3$                             &         $0$           &        $2$      \\
 4&  0     & 0      &   $3.42\times10^{-2}$      &    $1.2976$      &        $3$       \\
 5& 0.111  & -0.689 &    $2.83\times10^{-6}$     &     $1.8947$     &        $11$      \\
 6& 0.509  & 0.589  &     $1.8\times10^{-13}$     &    $1.9898$      &       $13$       \\
 7& -0.424 & 0.592  &                $0$                      &          $0$          &       $20$     \\
 \hline\hline
\end{tabular}
\end{table}

Now, we would like to find the energy density of the sample. To do so, let us consider $N_p$ particles interacting via the pairwise potential (\ref{Eq1}). The corresponding magnetic energy of $N_p$ particles reads
\begin{equation}\label{Eq4}
U_N=N_pU_0+\frac{1}{2}\underset{i\neq j}{\sum_{j,i}} U_{ij}^{(\mathrm{int})}.
\end{equation}
We should note here, that the same equation (\ref{LES}) holds for electrically polarizable particles. And for that case, the interaction potential is found to be non-additive~\cite{Bossis93}. Thus, the results obtained by means of PIA may not represent accurately the case of very dense systems. However, it is instructive to estimate the upper limit for the particles density where dipolar interaction contribution is prevailing. Following our previous study~\cite{Yaremchuk2020}, we consider the sample of ellipsoidal shape which is characterized by a constant demagnetizing field~\cite{Osborn, Maxwell1881}. Moreover, we consider the arrangement of particles, such that, the configuration surrounding the $i$-th particle is independent on the actual number $i$. When both conditions are satisfied, the expression that remains after summation over $j$, is independent of $i$ and can be evaluated easily. This results in the expression
\begin{equation}\label{Eq6}
		U_N= N_pU_0\bigg(1+3\beta_1 \phi  \frac{1}{4\pi c}\sum_{j\neq i} F(r_{ij}) \bigg) 
\end{equation}
where $\phi=cV_p$ is the volume fraction of the particles and $c=N_p/V$ is their number density. Here we introduced the short-hand functions $F(r_{ij})$ and $F_k(r_{ij})$:
\begin{equation}\label{Fij}
F(r_{ij})=\sum_{k=3}^7 F_k(r_{ij})
\end{equation}
and
\begin{equation}\label{Fk}
F_k(r_{ij})=\beta_1^{p_k-2} \bigg[     \frac{a^{k-3}A_k}{(r_{ij}-aB_k)^k}+\frac{a^{k-3}C_k}{(r_{ij}-aD_k)^k}\cos^2{\theta_{ij}}   \bigg].
\end{equation}
From Eqs.~(\ref{Eq6})-(\ref{Fk}), one can obtain the energy density $u_N=U_N/V$ as follows
\begin{equation}\label{UthroughF}
u_N=u_0\bigg( 1+3\beta_1\phi F  \bigg),
\end{equation}
where $u_0=-\frac{3}{2}\mu_0\beta_1\phi H_0^2$. The dimensionless factor
\begin{equation}\label{F}
F=\frac{1}{4\pi c}\sum\limits_{j\neq i}F(r_{ij})
\end{equation}
is determined by both the distribution of the particles and the initial shape of the sample. The similar expression for $u_N$ was obtained by us earlier~\cite{Yaremchuk2020} using the infinite series Eq.~(\ref{LES}) which involves evaluation of the factor $F_{\mathrm{LE}}=\sum_k f^{(k)}$. The explicit forms of the first few terms, $f^{(3)}$, $f^{(6)}$, $f^{(8)}$ and $f^{(9)}$, are known from the previous studies~\cite{Jeff} and will be used in the current study.

\emph{\textbf{Equilibrium strain}.}
The total energy density of the MSE under magnetic field is as follows
\begin{equation}\label{MagneticEnergy}
u_{\mathrm{tot}}=\frac{E\varepsilon^2}{2}+u_0\bigg(1+3\beta_1 \phi
F(\varepsilon)
\bigg),
\end{equation}
where the first term represents elastic contribution, and the second one corresponds to the magnetic energy density of the sample. The $\varepsilon=(l-l_0)/l$ is the dimensionless strain of the sample along the direction of applied magnetic field. The equilibrium strain $\varepsilon_{\mathrm{eq}}$ is the result of minimization of the energy density, $\frac{\partial}{\partial \varepsilon}u_{\mathrm{tot}}=0$, which yields
\begin{equation}\label{Strain}
\varepsilon_{\mathrm{eq}}=-\frac{3\beta_1\phi u_0}{E}\bigg[\frac{\partial}{\partial \varepsilon}
F(\varepsilon)
\bigg]_{\varepsilon_{\mathrm{eq}}}.
\end{equation}
Because $u_0\leq0$, see Eq.~(\ref{Eq2}), the prefactor in front of square brackets is always positive.
In the zero approximation for $\frac{\partial}{\partial \varepsilon}F(r_{ij}(\varepsilon))$ with respect to $\varepsilon$, the above equation has the following solution
\begin{equation}\label{StrainOne}
\varepsilon_{\mathrm{eq}}^{(1)}=\bigg.\varepsilon_{\mathrm{eq}}
\bigg|_{\varepsilon=0}.
\end{equation}
It can be seen that the function
\begin{equation}\label{K}
\bigg[\frac{\partial}{\partial \varepsilon}
%\frac{1}{4\pi c}\sum_{j\neq i} F(r_{ij}(\varepsilon))
F(\varepsilon)
\bigg]_{\varepsilon=0}=K
\end{equation}
determines the sign of the strain. The case $\varepsilon_{\mathrm{eq}}^{(1)}<0$ corresponds to the contraction, whereas $\varepsilon^{(1)}_{\mathrm{eq}}>0$ corresponds to the elongation of the sample along the direction of the external field $\bm{H}_0$.

Let us compare the obtained expression (\ref{Strain}) for the equilibrium strain with the analogous result obtained in Ref.~\cite{Ivaneyko14} using dipole-dipole interactions only. It was assumed that $\bm{m}||\bm{H}_0$ and the magnetic energy of a sample is derived in a self-consistent manner. As a result, the following expression was obtained in Ref.~\cite{Ivaneyko14}
\begin{equation}\label{SrtainIv}
\varepsilon_{\mathrm{eq}}=\frac{1}{2E}\mu_0\bigg[\frac{\phi^2H_0^2}{\big(\chi^{-1}+1/3-\phi f^{(3)}\big)^2}\frac{\partial}{\partial \varepsilon} f^{(3)} \bigg]_{\varepsilon_{\mathrm{eq}}}.
\end{equation}
Rewriting the result from Ref.~\cite{Ivaneyko14} in terms of our notations
\begin{equation}
u_0=-\frac{3}{2}\mu_0\beta_1\phi H^2_0; \quad \beta_1=\frac{\chi}{\chi+3},
\end{equation}
one obtains
\begin{equation}\label{StrainOld}
\varepsilon_{\mathrm{eq}}=-\frac{3\beta_1\phi u_0}{E}\bigg[\frac{1}{\big(1-3\beta_1\phi f^{(3)}\big)^2}\frac{\partial}{\partial \varepsilon} f^{(3)} \bigg]_{\varepsilon_{\mathrm{eq}}}.
\end{equation}
For isotropic particle distributions, including the cases of certain lattice arrangements, the factor $f^{(3)}$ is bounded from above by $1/3$~\cite{Ivaneyko14}, and $\beta_1\approx 1$ for carbonyl iron particles~\cite{Biller2014a}. Thus,
for small volume fraction of particles $\phi\ll 1$, one can obtain the following approximation
\begin{equation}\label{StrainOldSmall}
\varepsilon_{\mathrm{eq}} \approx -\frac{3\beta_1\phi u_0}{E}\bigg[\frac{\partial}{\partial \varepsilon} f^{(3)} \bigg]_{\varepsilon_{\mathrm{eq}}}.
\end{equation}
At the same conditions, inhomogeneous magnetization can be ignored and one obtains the following approximation
\begin{equation}\label{StrainApprox}
\frac{1}{4\pi c}\sum\limits_{j\neq i}  F(r_{ij}(\varepsilon))\approx f^{(3)}.
\end{equation}
Now, comparing the Eqs.~(\ref{Strain}, \ref{StrainApprox}) with Eq.~(\ref{StrainOldSmall}), one can conclude that both models predict the same result for the equilibrium strain in the limit of small volume fraction $\phi~\ll~1$.

\section{The influence of the initial shape of the sample on the magnetostriction effect}\label{III}

It was shown previously~\cite{Ivaneyko14}, that the summation $\frac{1}{4\pi c}\sum\limits_{j\neq i}$ over the particles positions inside the spheroidal sample can be split into two terms. First term involves the summation within the proximity region of the chosen particle (termed as a micro-sphere), where the result is highly dependent on the particles distribution in space. Second term is the one outside the micro-sphere, where exact particles positions are not important, and the integration over this spatial region can be undertaken instead of summation. Minimal radius of such a micro-sphere was estimated (for the uniform distribution) to be $R_{\mathrm{ms}}\sim 10\langle r_{n.n}\rangle$~\cite{Ivaneyko14}, where $\langle r_{n.n}\rangle$ is an average distance between the nearest neighbors. This formalism is convenient, since it allows a separate treatment of different effects. For example, the dimensionless factor $F(\varepsilon)$
can be split into two terms
\begin{flalign}
&F(\varepsilon)\!=\!\frac{1}{4\pi c}\sum_{r_{ij}<R_{\mathrm{ms}}} F(r_{ij}(\varepsilon))\!+\!\frac{1}{4\pi}\int_{V_{\mathrm{ms}}}^{V_{\mathrm{s.sample}}} dV  F(r(\varepsilon))=&\nonumber\\
&=F_{\mathrm{micro}}(\varepsilon)+F_{\mathrm{macro}}(\varepsilon),&
\label{FmFm}\end{flalign}
where integration is performed over the entire volume of a sample save the micro-sphere. The first term provides the contribution to the magnetic energy from the particles distribution, whereas the second one accounts for the role of the sample shape. The effect on the magnetostriction can be treated similarly. The function $K$ from the Eq.~(\ref{K}) can be split as follows~\cite{Dirk19}
\begin{equation}\label{KmicroKmacro}
\! K\!\!=\!\!\bigg[\frac{\partial}{\partial \varepsilon}F_{\mathrm{micro}}(\varepsilon)\bigg]_{\varepsilon=0}\!\!+\!\!\bigg[\frac{\partial}{\partial \varepsilon}F_{\mathrm{macro}}(\varepsilon) \bigg]_{\varepsilon=0}\!\!\!=\!\!K_{\mathrm{micro}}+K_{\mathrm{macro}},
\end{equation}
where $K_{\mathrm{micro}}$ and $K_{\mathrm{macro}}$ represent the same respective contributions as in Eq.~(\ref{FmFm}).

\emph{\textbf{Shape factor}.}
Let us consider now the contribution related to the initial shape in more detail. To do so, we repeat here the procedure outlined in Ref.~\cite{Ivaneyko14}. Assuming that the micro-sphere is situated in the center of the spheroidal sample, the integration in Eq.~(\ref{FmFm}) is greatly simplified
\begin{equation}
\frac{1}{4\pi}\int_{V_{\mathrm{ms}}}^{V_{\mathrm{s.sample}}} dV=\frac{1}{4\pi}\int_0^{2\pi}d\varphi \int_0^{\pi}d\theta \sin{\theta}\int_{R_{\mathrm{ms}}}^{r_{\theta}}dr r^2,
\end{equation}
where $r_{\theta}=A\big[(1-\gamma^2_\varepsilon)\cos^2{\theta}+\gamma^2_\varepsilon\big]^{-\frac{1}{2}}$ is the upper bound of integration over $r$. Parameter $\gamma_\varepsilon=A/B$ is the aspect ratio of a spheroid with the semi- axes $A$ and $B$ which explicitly depends on the sample strain $\varepsilon$. According to Table~\ref{tab:table1}, the first term in Eq.~(\ref{Fk}) is
\begin{equation}
F_3(r_{ij}(\varepsilon))= \frac{1}{r_{ij}^3(\varepsilon)}\bigg[3\cos^2{\theta_{ij}}(\varepsilon)-1\bigg].
\end{equation}
It is a dipolar term, and after performing the integration one obtains
\begin{equation}\label{F3}
F^{(3)}_{\mathrm{macro}}(\varepsilon)=\frac{1}{3}-N(\gamma_\varepsilon),
\end{equation}
where $N(\gamma_\varepsilon)$ is a well known demagnetizing factor~\cite{Dirk19, Osborn}:
\begin{equation}\label{DeffN}
N(\gamma_\varepsilon)=\frac{\gamma_\varepsilon}{(\gamma_\varepsilon^2-1)^{3/2}}\ln{\big(\gamma_\varepsilon+\sqrt{\gamma_\varepsilon^2-1}\big)}-\frac{1}{\gamma_\varepsilon^2-1}.
\end{equation}
The demagnetizing factor, $N(\gamma_\varepsilon)$, is related to the external surface of the entire sample, whereas $1/3$ term in Eq.~(\ref{F3}) is its analogue, related to the micro-sphere.
Let us note that the later term will cancel out with its counterpart originated from summation over the micro-sphere.
It is clear, that $F^{(3)}_{\mathrm{macro}}(\varepsilon)$ depends on the aspect ratio $\gamma_\varepsilon$ only and is dimensionless. Contrary to that, the other terms, $F^{(k>3)}_{\mathrm{macro}}(\varepsilon)$, contain the radius of the particle, $a$, in their respective expressions, which is microscopic lengths scale. In general, for some $k>3$ one can write
\begin{equation}
F_{\mathrm{macro}}^{(k>3)}=\beta_1^{p_k-2}\!\!\int_0^1\!\!\! dx \int_{R_{\mathrm{ms}}}^{r_x}\!\!\! dr r^2\bigg[     \frac{a^{k-3}A_k}{(r\!-\!aB_k)^k}+\frac{a^{k-3}C_k}{(r\!-\!aD_k)^k}x^2   \bigg],
\end{equation}
where  $r_x=A\big[(1-\gamma^2_\varepsilon)x^2+\gamma^2_\varepsilon\big]^{-\frac{1}{2}}$ is proportional to the $A$ semi-axis of a spheroid. This integration leads to the negligible contributions to $F_{\mathrm{macro}}(\varepsilon)$ (see Eq.~(\ref{AppIntegral})). As the result, for macroscopic sample one can use Eq.~(\ref{F3})
\begin{equation}\label{FmacroFinal}
F_{\mathrm{macro}}(\varepsilon)=\frac{1}{3}-N(\gamma_\varepsilon).
\end{equation}
In other words, due to the short-range nature of the interactions in $F^{(k>3)}$, the shape factor is determined solely by a dipolar potential.

\emph{\textbf{Shape-dependent magnetostriction}.}
Now, let us turn our attention to the $K_{\mathrm{macro}}$. First of all, we need to specify the dependency of the aspect ratio $\gamma_\varepsilon$ on the strain $\varepsilon$. If one assumes the linear affine deformation with respect to a chosen axis, then the requirement of the volume conservation yields $\gamma_\varepsilon=\gamma_0(1+\varepsilon)^{3/2}$. Thus, the derivative over strain is $\big[\frac{\partial}{\partial \varepsilon}F_{\mathrm{macro}}(\varepsilon)\big]_{\varepsilon=0}=\frac{3}{2}\gamma_0\frac{\partial}{\partial \gamma_0}F_{\mathrm{macro}}(\gamma_0)$. One can check (see Eq.~(\ref{AppKFinal})), that when calculating $K_{\mathrm{macro}}$, the Eq.~(\ref{FmacroFinal}) can be used. Following the Eq.~(\ref{DeffN}), one can obtain $K_{\mathrm{macro}}(\gamma_0)=\frac{\partial}{\partial \varepsilon}F_{\mathrm{macro}}(\varepsilon=0)$ in the following form
\begin{equation}
K_{\mathrm{macro}}(\gamma_0)=\frac{3}{2}\frac{2\gamma_0^3+\gamma_0}{(\gamma_0^2\!-\!1)^{5/2}}\ln{\big[\gamma_0+\sqrt{\gamma_0^2\!-1\!}\big]}-\frac{3}{2}\frac{3\gamma_0^2}{(\gamma_0^2\!-\!1)^2}.
\end{equation}
This expression was obtained in Ref.~\cite{Dirk19} previously for the case of dipolar interactions. As we have shown, it holds for a more general case as well. It can be seen that the shape dependent magnetostriction factor, $K_{\mathrm{macro}}$, is always positive, what means the expansion of the sample along the magnetic field.

In the current chapter we demonstrated the agreement of our results with the ones of the continuum mechanics theory~\cite{Landau,Maxwell1881, Osborn}, namely, that the shape dependent magnetostriction is determined solely by the dipolar interactions.

\section{The influence of the spatial distribution of magnetic particles on the magnetostriction effect}\label{IV}

In this section we will discuss the contribution to the magnetostriction which originates from spatial distribution of magnetic particles, given by the $K_{\mathrm{micro}}$ term in Eq.~(\ref{KmicroKmacro}). We assume the case of strong coupling between magnetic particles and a host polymeric matrix, as in case of magnetic particles embedded into a polymer via covalent bonds. In this case, application of the external magnetic field results in an affine deformation of a sample. The case of weak coupling is also possible when magnetic particles are just physically embedded inside a soft elastic matrix~\cite{Odenbach2014} and move independently of the matrix. In our case, the effect of different spatial distributions on the micro-sphere magnetostriction factor, $K^{\mathrm{affine}}_{\mathrm{micro}}$, is exemplified by the uniform and four lattice-like particle arrangements.

\subsection{Approach based on the approximation of the solution for the Laplace equation}

To this end we evaluate the function $F_k(r_{ij})$ and the first few terms $f^{(k)}$, all originated from the interaction energy Eq.~(\ref{LES}). Let us consider the former case in more detail here. The linear affine deformation along the $Ox$ axis~\cite{Dirk19}, and the volume conservation requirement would lead to the following relations: $r_x=r^{(0)}_x(1+\varepsilon)$, $ r_y=r^{(0)}_y\frac{1}{\sqrt{1+\varepsilon}}$ and $r_z=r^{(0)}_z\frac{1}{\sqrt{1+\varepsilon}}$ between the initial $(r^{(0)}_x, r^{(0)}_y, r^{(0)}_z)$ and after-the- deformation $(r_x, r_y, r_z)$ positions of each particle. The $F_{\mathrm{micro}}(\varepsilon)$ is, according to Eq.~(\ref{FmFm}), a sum over the micro-sphere as follows
\begin{equation}
F_{\mathrm{micro}}(\varepsilon)=\frac{1}{4\pi c}\sum_{r_{ij}<R_{\mathrm{ms}}}\sum_{k=3}^7 F_k(r_{ij}(\varepsilon)).
\end{equation}
Using the explicit form for the $F(r_{ij}(\varepsilon))$ from Eqs.~(\ref{Fij},\ref{Fk}), and rewriting $\cos^2{\theta_{ij}}$ as the fraction $\frac{(r_x)^2_{ij}}{r^2_{ij}}$, one can obtain the following
	\begin{flalign}
	&	\bigg[\frac{\partial}{\partial \varepsilon}F_{\mathrm{micro}}(\varepsilon)\bigg]_{\varepsilon=0}=\frac{1}{\phi}\!\!\sum_{r_{ij}<R_{\mathrm{ms}}}\sum_{k=3}^7 \bigg[g_{4,r^{(0)}_{ij}}^{(k)}\cos^4{\theta_{ij}}+&\nonumber\\
&+g_{2,r^{(0)}_{ij}}^{(k)}\cos^2{\theta_{ij}}+g_{0,r^{(0)}_{ij}}^{(k)}\bigg],&
 \label{FulldFk}   \end{flalign}
where we used the relation between volume fraction of particles, $\phi$, and their number density, $c$, in the form of $\frac{3}{4\pi c a^3}=1/\phi$. The functions $g^{(k)}_{m,r^{(0)}_{ij}}$ are, in general, dependent on the dimensionless inverse distance $a/r^{(0)}_{ij}$. With the help of some ancillary notation,
\begin{equation}\label{helpful} g_s(P;R)=\big(\frac{a}{r^{(0)}_{ij}}\big)^{k}\frac{\beta_1^{p_k-2}P}{(1-\frac{a}{r^{(0)}_{ij}}R)^k}\big(\frac{k}{s}\frac{1}{1-\frac{a}{r^{(0)}_{ij}}R}+1\big),
\end{equation}
these functions can be written in relatively compact form as follows
\begin{equation}\label{g_4-g_0}
	\begin{aligned}
   \!\!\! g^{(k)}_{4,r^{(0)}_{ij}}\!=\!&-g_2(C_k;D_k), \quad g^{(k)}_{0,r^{(0)}_{ij}}=g_6(A_k;B_k),\\
		\!\!\!\!&g^{(k)}_{2,r^{(0)}_{ij}}\!\!=\! g_6(C_k;D_k)-g_2(A_k;B_k).
	\end{aligned}	
\end{equation}
The constants $A_k,\dots,D_k$ can be found in the Table.~(\ref{tab:table1}).

One should note, however, that in this affine deformation formalism we implicitly take into account the particles that are located outside the micro-sphere after the deformation, or were inside it before the deformation occurs. We should exclude the effect of the boundary crossing, denoted by $C_{\mathrm{b.c}}$, from the final result. More details can be found in~\cite{Dirk19} as well as in Eq.~(\ref{AppBorderCrossingDipole}). As the result, we obtain the following expression
\begin{equation}\label{Kaffinemicro}
K_{\mathrm{micro}}^{\mathrm{affine}}= \bigg[\frac{\partial}{\partial \varepsilon}F_{\mathrm{micro}}(\varepsilon)\bigg]_{\varepsilon=0}-\frac{2}{5}.
\end{equation}

Let us consider the contribution to $K_{\mathrm{micro}}^{\mathrm{affine}}$ from the $k=3$ term in Eq.~(\ref{FulldFk}). In this case, the coefficients $g_{m,r^{(0)}_{ij}}^{(3)}$ are greatly simplified and  one can obtain the following
\begin{equation}\label{KaffinemicroThreeThroughg}
\!\!\! K_{\mathrm{micro}}^{\mathrm{affine}}(3)=\!\frac{1}{\phi}\!\!\!\sum_{r^{(0)}_{ij}<R_{\mathrm{ms}}} \!\!\!\! \frac{ -15\cos^4{\theta_{ij}}+12\cos^2{\theta_{ij}}-1}{2(r^{(0)}_{ij})^3}-\frac{2}{5},
\end{equation}
which coincides with the expression for $K_{\mathrm{micro}}^{\mathrm{affine}}$ obtained for the dipole-dipole interaction in Ref.~\cite{Dirk19}. Other terms, which reflect the effect of the inhomogeneous magnetization of the particles can be expressed as follows
	\begin{flalign}
	&K^{\mathrm{affine}}_{\mathrm{micro}}(k>3)=\frac{1}{\phi}\sum_{k=4}^7\bigg[g_{4,r^{(0)}_{ij}}^{(k)}\cos^4{\theta_{ij}}+g_{2,r^{(0)}_{ij}}^{(k)}\cos^2{\theta_{ij}}+&\nonumber\\
&+g_{0,r^{(0)}_{ij}}^{(k)}\bigg].&
\label{KRestMain}
\end{flalign}

As a next step, let us consider the sign of the contribution $K_{\mathrm{micro}}^{\mathrm{affine}}$ to the magnetostriction factor for different particle distributions.

\emph{\textbf{Uniform arrangement of magnetic particles}.}
Micro-sphere is centered around the randomly chosen $i$-th particle and the pair distances are $R_{\mathrm{ms}}\geq r^{(0)}_{ij}\geq r^{(0)}_{\mathrm{1s}}$, where $r^{(0)}_{\mathrm{1s}}$ defines the first coordination sphere. Because of the finite size of the particles and related to it excluded volume effects, the genuinely uniform (UNF) distribution can be hardly achieved here. However, we can assume such distribution on average, namely by substitution the sum as follows: $\sum_{j,i} U_{ij}^{(\mathrm{int})}=N_p\sum_j\langle U_{ij}^{(\mathrm{int})}\rangle$. Conceptually similar smearing off for the particle positions is reported in Refs.~\cite{Dirk, Dirk2022}. Thus, we replace the summation in Eq.~(\ref{Kaffinemicro}) by the integration $\frac{1}{c}\sum\rightarrow\int dV$. The positions of the particles inside integration domain are not correlated with the absolute value of the radius-vector under uniform distribution assumption, and the integration over the distance and the solid angle are independent. Applying the expressions $\frac{1}{4\pi}\int d\Omega \cos^2{\theta}=1/3$ and $\frac{1}{4\pi}\int d\Omega \cos^4{\theta}=1/5$ to the Eq.~(\ref{KaffinemicroThreeThroughg}), one can see that $K_{\mathrm{micro}}^{\mathrm{affine}}$ turns into a constant
\begin{equation}\label{KaffinemicroThreeIsotrAverageMain}
K_{\mathrm{micro}}^{\mathrm{affine}}(3)|_{\mathrm{UNF}}=-\frac{2}{5},
\end{equation}
which is the known result, obtained earlier in Ref.~\cite{Dirk19}.
To calculate the other terms $K_{\mathrm{micro}}^{\mathrm{affine}}(k>3)|_{\mathrm{UNF}}$ in Eq.~(\ref{KRestMain}), we use the Eqs.~(\ref{g_4-g_0},\ref{helpful}) and, as the result, obtain the following expression
\begin{equation}\label{KIsotrkIntegralMain}
\!\!\! K^{\mathrm{affine}}_{\mathrm{micro}}(k \! >\!3)\big|_{\mathrm{UNF}}\!=\!\sum_{k=4}^7\!\!\frac{2\beta_1^{p_k-2}}{15}\frac{C_k(y+D_k)^3}{y^k}\bigg|^{R_{\mathrm{ms}}/a-D_k}_{r^{(0)}_{\mathrm{1s}}/a-D_k}\!\!,
\end{equation}
where the constants $C_k$, $D_k$ and $p_k$ can be found in the Table~\ref{tab:table1}, and the change of the variable $y=r^{(0)}_{ij}/a-D_k$ was done during the integration. The result in Eq.~(\ref{KIsotrkIntegralMain}) does not depend on the constants $A_k$ and $B_k$ because the corresponding terms cancel out when $\frac{1}{4\pi}\int d\Omega \cos^2{\theta}=1/3$.

\begin{figure}[h!]
\centering
  \includegraphics[scale=0.69]{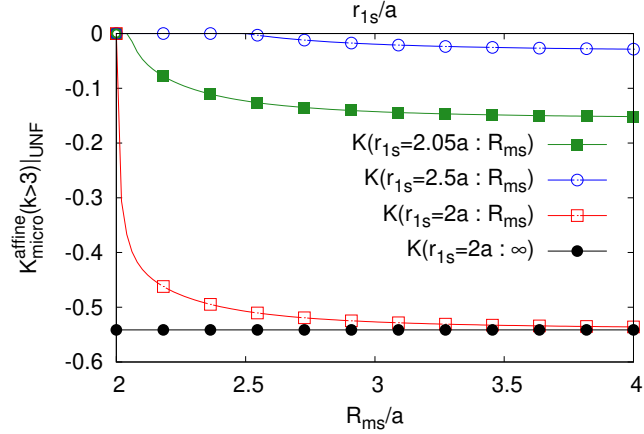}
\caption{Dependencies of the function $K^{\mathrm{affine}}_{\mathrm{micro}}(k>3)\big|_{\mathrm{UNF}}$ from Eq.~(\ref{KIsotrkIntegralMain}) on various integration boundaries. In all cases the function is negative. If lower bound is small, $r^{(0)}_{\mathrm{1s}}=2a$, the function converges to the value $\sim-0.54$ already for the micro-sphere radius of $R_{\mathrm{ms}}\sim 3a$. Also, one can observe that the magnitude of the function diminishes considerably even for small increment of the loser bound of the integration $r^{(0)}_{\mathrm{1s}}$.}
 \label{fig:KIsotrkIntMain}
\end{figure}

The dependence of this function on the lower and upper integration bounds is plotted in Fig.~(\ref{fig:KIsotrkIntMain}). It is clear that the function is negative for all the cases. It converges rather quickly to a finite value when the radius of the micro-sphere $R_{\mathrm{ms}}/a$ gets larger. For example, it converges to the value $\sim-0.54$ for the micro-sphere of the radius $R_{\mathrm{ms}}\sim 3a$ when the lower bound of the integral is equal to $r^{(0)}_{\mathrm{1s}}=2a$. As the lower integration bound increases, the function converges to the lower value as well, and for $r^{(0)}_{\mathrm{1s}}=2.5a$ the contribution from the $k>3$ terms is significantly smaller than the dipole $k=3$ contribution, as given by Eq.~(\ref{KaffinemicroThreeIsotrAverageMain}). And, being the constant, the dipole term only shifts all graphs down, and do not change their dependence on distances.

\emph{\textbf{Lattice-like arrangement of magnetic particles}.}
The lattice-like arrangements of particles presents great theoretical and experimental interest. The presence of the well-defined spatial patterns of the particles positions in this case helps to link them with the properties of the MSE. Such arrangements can be achieved experimentally. In particular, Zhang et al.~\cite{Zhang2008} reported the process of preparation the sample with arbitrary lattice-like arrangement of the particles that are 400-1000~{\textmu}m in diameter. Quick creation and transfer into different substrates of $2$D patterns with smaller metallic particles was reported as well~\cite{Zhang2019}, though without mentioning of MSE preparation.

In order to analyze the influence of the different distributions of the particles on the magnetostriction effect, it may be convenient to rearrange the sum, like the one in Eq.~(\ref{Kaffinemicro}), into the series of sums performed over the coordination spheres ($1s, 2s,\ldots$). For instance, the summation of the $\cos^q{\theta_{ij}}$ term can be rearranged in the following way
\begin{equation}\label{CoordSpherSumMain}
\sum_{r^{(0)}_{ij}<R_{\mathrm{ms}}} \cos^q{\theta_{ij}}=\sum_{r^{(0)}_{ij}\in 1s}\cos^q{\theta_{ij}}+\sum_{r^{(0)}_{ij}\in 2s}\cos^q{\theta_{ij}}+\cdots
\end{equation}
where the summation indices, $r^{(0)}_{ij}\in ns$, denote summation over the $n$-th coordination sphere. In this way one can separate the angular and radial dependent parts, because in each $n$-th coordination sphere the radius vectors $r^{(0)}_{ij}$ are constant, $r^{(0)}_{ns}$. Additionally, introducing the $N_{\mathrm{ns}}$ as the number of particles on the $n$-th coordination sphere, one can define the average of some angular dependent function
$\langle \cos^q{\theta_{ij}} \rangle_{\mathrm{ns}}$ as follows
\begin{equation}\label{CoordSpherAverMain}
\langle\cos^q{\theta_{ij}}\rangle_{\mathrm{ns}}=\frac{1}{N_{\mathrm{ns}}}\sum_{r^{(0)}_{ij}\in ns}\cos^q{\theta_{ij}},
\end{equation}
which is simply the arithmetic mean of such function on the $n$-th coordination sphere. In general, the average will depend on the particular spatial distribution of particles.

\begin{figure}
\centering
\includegraphics[width=.99\linewidth]{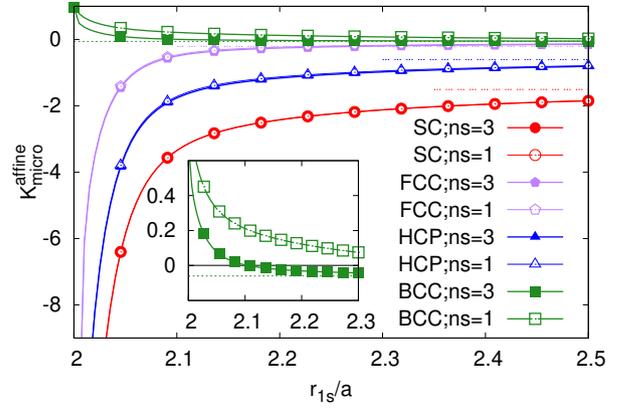}
  \caption{Dependency of the function $K^{\mathrm{affine}}_{\mathrm{micro}}\big|_{\mathrm{distr.}}$ given by Eq.~(\ref{TogetherEq}), on the radius of the first coordination sphere $r^{(0)}_{\mathrm{1s}}\in[2.02a,2.52a]$. It is calculated for simple cubic (SC), body-centered cubic (BCC), hexagonal close-packed (HCP) and face-centered cubic (FCC) lattices. Its asymptotic values, given by the dipolar contribution including the border crossing term, $2/5$,~\cite{Dirk19}, are shown as dashed lines. The inset provides a magnified view for the BCC lattice results at $r^{(0)}_{\mathrm{1s}}\approx2a$. For this case, the contribution to the magnetostriction factor from a micro-sphere changes its sign at around $r^{(0)}_{\mathrm{1s}}\approx2.1a$.}

  \label{fig:DistributionsMain}
\end{figure}

In the case of the dipolar interparticle interaction $(k~=~3)$, the magnetostriction factor $K_{\mathrm{micro}}^{\mathrm{affine}}(k=3)$ was already calculated in Ref.~\cite{Dirk19} for the cases of the simple cubic (SC), body-centered cubic (BCC) and hexagonal close-packed (HCP) lattices inside the micro-sphere of relatively large size $R_{\mathrm{ms}}$. We complemented these findings by the case of the FCC lattice, where the magnetic field, $\bm{H}_0$, was chosen to be collinear to one of the edges of a cube. The micro-sphere of the radius $R_{\mathrm{ms}}=100r_{1s}$ was considered.

The functions $g^{(k)}_{m,r^{(0)}_{ij}}$, entering Eqs.~(\ref{FulldFk},\ref{KRestMain}) and defined via  Eqs.~(\ref{g_4-g_0},\ref{helpful}), alongside with the functions
	\begin{flalign}
	&M_k^\mathrm{distr.}(r^{(0)}_{ns})=\big\langle g^{(k)}_{4,r^{(0)}_{ns}}\cos^4{\theta_{ij}}+g^{(k)}_{2,r^{(0)}_{ns}}\cos^2{\theta_{ij}}+\nonumber\\&+g^{(k)}_{0,r^{(0)}_{ns}}\big\rangle_\mathrm{ns}^\mathrm{distr.},
	\label{Mkdistr}\end{flalign}
can be used to define the contribution from the beyond dipolar interactions at each coordination sphere as follows
\begin{equation}\label{MkDistrMain}
M_{k>3}^{\mathrm{distr.}}(r^{(0)}_{\mathrm{ns}})=\sum_{k=4}^{7}M^{\mathrm{distr.}}_k(r^{(0)}_{\mathrm{ns}}).
\end{equation}
This allows us to write the respective contributions to the magnetostriction factor
\begin{flalign}
&K^{\mathrm{affine}}_{\mathrm{micro}}(k>3)\big|_{\mathrm{distr.}}=\frac{1}{\phi}\big[N_{\mathrm{1s}}M_{k>3}^{\mathrm{distr.}}(r^{(0)}_{\mathrm{1s}})+\nonumber\\&+N_{\mathrm{2s}}M_{k>3}^{\mathrm{distr.}}(r^{(0)}_{\mathrm{2s}})+\cdots\big].
\label{KmicroLatticesDeffMain}\end{flalign}
Using Table~\ref{tab:table5} one can calculate $K^{\mathrm{affine}}_{\mathrm{micro}}(k>3)$ given by the Eq.~(\ref{KmicroLatticesDeffMain}) for different lattice types.

The results discussed above allow us to consider the complete expression for the magnetostriction factor inside the micro-sphere
\begin{equation}\label{TogetherEq}	K_{\mathrm{micro}}^{\mathrm{affine}}=K^{\mathrm{affine}}_{\mathrm{micro}}(k=3)+K^{\mathrm{affine}}_{\mathrm{micro}}(k>3).
\end{equation}
The corresponding plots are shown in the Fig.~(\ref{fig:DistributionsMain}) for the interval $r^{(0)}_{\mathrm{1s}}=2.02a-2.52a$. It can be seen, that in all cases the magnitude of this function decreases with the increase of the radius of the first coordination sphere $r^{(0)}_{\mathrm{1s}}$. This reflects a short-range nature of the interactions leading to the inhomogeneous magnetization effects. Similarly, it will decrease as the volume fraction of the particles, $\phi$, decreases. For the SC, FCC, and HCP lattices, this factor is always negative, indicating the contribution from a micro-sphere towards the sample contraction along the field. Contrary to that, $K_{\mathrm{micro}}^{\mathrm{affine}}$ changes sign from positive to negative at around $r^{(0)}_{\mathrm{1s}}\approx2.1a$ for the BCC lattice case. Beside this feature, in the case of BCC lattice, the contributions from the first and from the following coordination spheres are visibly different (see inset in Fig.~(\ref{fig:DistributionsMain})). We explain this by the proximity of the first, $r^{(0)}_{\mathrm{1s}}$, and of the second, $r^{(0)}_{\mathrm{2s}}$, coordination spheres in the BCC lattice: $r^{(0)}_{\mathrm{2s}}=\sqrt{4/3}r^{(0)}_{\mathrm{1s}}$.
Because their respective contributions to the magnetostriction factor, $K_{\mathrm{micro}}^{\mathrm{affine}}$, are of the opposite sign, they compete strongly in a narrow range of distances.

The dipolar contributions are represented by their respective asymptotic values, $-1.51$, $-0.06$, $-0.21$ and $-0.10$, for the case of SC, BCC, HCP and FCC lattices, shown as dashed lines in Fig.~(\ref{fig:DistributionsMain}). One can see that they approximate the function $K_{\mathrm{micro}}^{\mathrm{affine}}$ adequately when the nearest neighbors distance reaches certain characteristic value. This is discussed in more detail latter, when comparing magnitude of dipolar and higher contributions to the micro-sphere magnetization factor $K_{\mathrm{micro}}^{\mathrm{affine}}$.

\begin{figure}
	\centering
	\includegraphics[width=.9\linewidth]{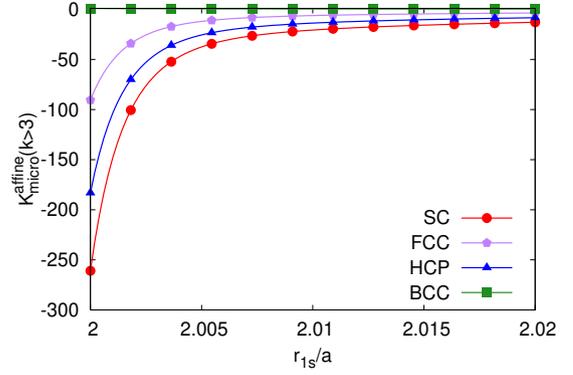}
	\caption{Dependence of the higher than dipolar magnetostriction factor, $K^\mathrm{affine}_\mathrm{micro}(k>3)$, given by the Eq.~(\ref{KmicroLatticesDeffMain}), on the radius of the first coordination sphere in the interval $r^{(0)}_{1s}\in[2a;2.02a]$.}
	\label{fig:DistributionsContact}
\end{figure}

At the interparticle separation equal to the first coordination sphere radius, $r^{(0)}_{\mathrm{1s}}=2a$, the function $K^{\mathrm{affine}}_{\mathrm{micro}}$, except for the case of the BCC lattice, acquires gigantic negative values (see the Fig.~(\ref{fig:DistributionsContact})). Despite the fact that the forces between inhomogeneously magnetized particles can be much larger than the dipolar ones~\cite{Biller2014}, we expect that this prediction for magnetostriction factor is unphysical. This, in part, is due to the fact, that in our model we cut off the interaction potential at the contact distances and do not introduce the short-range elastic repulsion~\cite{Goh2022}. Accounting for such repulsion terms may correct the model prediction at small interparticle distances. Moreover, at this close separations between particles, one expects a deviation from the affine deformation, as well as the increased significance of the three- and higher particle interactions.

\begin{figure}[h!]
\centering
 \includegraphics[width=.95\linewidth]{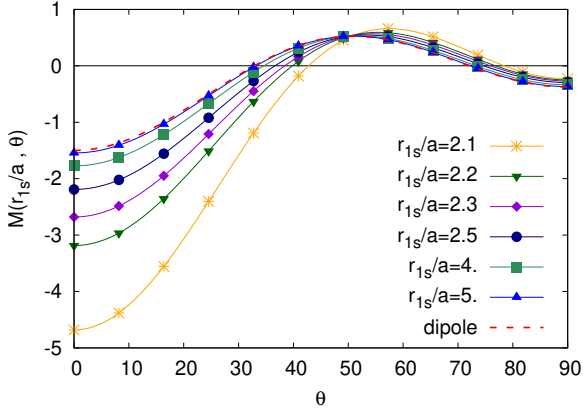}
  \caption{Dependencies of the functions $M(r^{(0)}_{\mathrm{1s}}; \theta)$ on the angle $\theta$ for various values of the first coordination sphere radius $r^{(0)}_{\mathrm{1s}}/a$. As distance between particles become smaller, the differences of the magnitudes become more apparent. The effect is more pronounced for the small angles $\theta\sim 0^{\circ}$, which corresponds to the chain-like conformation of the particles.}\label{fig:MAngularMain}
\end{figure}

As was discussed above, the behavior of the micro-sphere magnetostriction factor, $K^{\mathrm{affine}}_{\mathrm{micro}}$, exhibits a strong dependence on the lattice symmetry. This is especially true at the distances comparable to the first coordination sphere, and the contributions to $K^{\mathrm{affine}}_{\mathrm{micro}}(k>3)$ provided by the interactions on this lengths scale, could be sufficient for the description of their qualitative behavior. Concerning the dipolar term in Eq.~(\ref{TogetherEq}), its properties also can be inferred from the first coordination sphere contribution~\cite{Dirk19}. As can be seen in Eq.~(\ref{KmicroLatticesDeffMain}), the contribution to the magnetostriction factor from the nearest neighbors, ($r^{(0)}_{ij}=r^{(0)}_{\mathrm{1s}}$), is determined by the functions $M^{\mathrm{distr.}}_k(r^{(0)}_{\mathrm{1s}})$ given by the Eq.~(\ref{Mkdistr}). Therefore, let us introduce functions $M(r^{(0)}_{\mathrm{1s}}; \theta)$ defined via relations
\begin{equation}
	M(r^{(0)}_{\mathrm{1s}}; \theta)=\sum_{k=3}^7 M_k(r^{(0)}_{\mathrm{1s}}; \theta),
\end{equation}
\begin{equation}\label{MAngularDeffMain}
	\langle M_k(r^{(0)}_{\mathrm{1s}}; \theta) \rangle^{\mathrm{distr.}}_{\mathrm{1s}}=\frac{3}{4}(r^{(0)}_{\mathrm{1s}}/a)^3M^{\mathrm{distr.}}_k(r^{(0)}_{\mathrm{1s}}).
\end{equation}
The angle $\theta$ is defined in Eq.~(\ref{LES}) above.
The prefactor $(r^{(0)}_{\mathrm{1s}}/a)^3$ is used in order to account for the volume fraction of the particles, $\phi\propto (a/r^{(0)}_{\mathrm{1s}})^3$, entering the Eq.~(\ref{KmicroLatticesDeffMain}). This definition is convenient for the purpose of the direct comparison with the earlier results. Namely, using the information from the Table~\ref{tab:table1} and Eq.~(\ref{Mkdistr}), the dipolar contribution can be written explicitly as follows
\begin{equation}\label{DipoleAngularMain}
M_3(r^{(0)}_{\mathrm{1s}}; \theta)=\frac{3}{8}\big(-15\cos^4{\theta}+12\cos^2{\theta}-1\big).
\end{equation}
%%%%%%%%%%%%%%%%%%
This function is studied in Ref.~\cite{Dirk19}, and now we can compare it to the other contributions, arising from the inhomogeneous bulk magnetization of the particles. The dipolar term, $M_3(r^{(0)}_{\mathrm{1s}}; \theta)$, which is independent on the radius of the first coordination sphere, $r^{(0)}_{\mathrm{1s}}$, is showed as a dashed line in the Fig.~(\ref{fig:MAngularMain}). The function $M(r^{(0)}_{\mathrm{1s}}; \theta)$ start to deviate from the dipolar one when the distance $r_{\mathrm{1s}}$ is less than five particle radii. This effect is more pronounced for the chain-like ($\theta\sim 0^{\circ}$) arrangement of the particles. Qualitatively, the chain-like ($\theta\sim 0^{\circ}$) and plane-like ($\theta\sim 90^{\circ}$) conformations correspond to the negative values of the function $M(r^{(0)}_{\mathrm{1s}}; \theta)$. For the intermediate particle conformation the interval exists where the function is positive. In the case of the dipolar term $M_3(r^{(0)}_{\mathrm{1s}}; \theta)$, this region can be identified as $\theta\sim 40^{\circ}-70^{\circ}$. For the combined function $M(r^{(0)}_{\mathrm{1s}}; \theta)$ the interval where it is positive can slightly vary with increasing radius of the first coordination sphere, exhibiting shift towards the larger values of the angle $\theta$.

\subsection{Approach based on the accounting for explicit terms in the solution of the Laplace equation}
As was briefly discussed in Sec.~\ref{II}, the magnetic energy density of the sample can be written in the following form
\begin{equation}\label{AlternativeU}
	u_N=u_0\big( 1+3\beta_1\phi F_{\mathrm{LE}}  \big),
\end{equation}
where $F_{\mathrm{LE}}$ was obtained from the Eq.~(\ref{LES}) and has the form of infinite series $F_{\mathrm{LE}}=\sum_k f^{(k)}$ of dimensionless terms $f^{(k)}$~\cite{Yaremchuk2020}. The explicit expressions for the first four non-zero terms $f^{(k)}$, entering the expression for $F_{\mathrm{LE}}$, are known~\cite{Jeff}. This allows us to derive the micro-sphere magnetostriction factor $K^{\mathrm{affine}}_{\mathrm{micro}}$ in an alternative way to the one based on Eq.~(\ref{UthroughF}) and considered in this paper above.

In a way similar to Eq.~(\ref{FmFm}), the dimensionless terms $f^{(k)}$
each can be split into two parts, namely $f^{(k)}_{\mathrm{micro}}$, and $f^{(k)}_{\mathrm{macro}}$. The ``macro'' contribution is discussed in more detail in Ref.~\cite{Yaremchuk2020}. In this section we consider the ``micro'' terms. The first four of them are as follows~\cite{Jeff,Yaremchuk2020}
\begin{equation}\label{flist1}
f_{\mathrm{micro}}^{(3)}=\frac{1}{4\pi c}\sum_{r_{ij}<R_{\mathrm{ms}}} \frac{3(r_{ij})_x^2-r^2_{ij}}{ r_{ij}^5},
\end{equation}
\begin{equation}
 f_{\mathrm{micro}}^{(6)}=\frac{\beta_1 a^3}{4\pi c}\sum_{r_{ij}<R_{\mathrm{ms}}} \frac{3(r_{ij})_x^2+r^2_{ij}}{r_{ij}^8},
\end{equation}
\begin{equation}
f_{\mathrm{micro}}^{(8)}=\frac{3\beta_2 a^5}{4\pi c}\sum_{r_{ij}< R_{\mathrm{ms}}} \frac{2(r_{ij})_x^2+r^2_{ij}}{r_{ij}^{10}},
\end{equation}
\begin{equation}\label{flist4}
 f_{\mathrm{micro}}^{(9)}=\frac{\beta^2_1 a^6}{4\pi c}\sum_{r_{ij}< R_{\mathrm{ms}}} \frac{9(r_{ij})_x^2-r^2_{ij}}{r_{ij}^{11}}.
\end{equation}
Assuming the affine deformation, as it was the case in the previous sections, one can examine the respective magnetostriction factor $K_{\mathrm{micro}}^{\mathrm{affine}}(k)$ for each of these terms.
As expected, the case of $k=3$ corresponds to a dipolar contribution
\begin{equation}
K_{\mathrm{micro}}^{\mathrm{affine}}(3)=\bigg[\frac{\partial}{\partial \varepsilon} f_{\mathrm{micro}}^{(3)}(\varepsilon) \bigg]_{\varepsilon=0}-C^{(3)}_{\mathrm{b.c}}=
\end{equation}
\begin{equation}\label{fKmicrodipole}
\!\!\!=\frac{1}{4\pi c}\!\!\!\sum_{r^{(0)}_{ij}<R_{\mathrm{ms}}}\!\!\!\frac{1}{2( r^{(0)}_{ij})^3}\bigg[-45\cos^4{\theta_{ij}}+36\cos^2{\theta_{ij}}-3\bigg]-\frac{2}{5},
\end{equation}
where value $C^{(3)}_{\mathrm{b.c}}=2/5$ of border crossing term~\cite{Dirk19} is used. In the case of $k>3$, the border crossing terms are proportional to the negligibly small fractions $\big(\frac{a}{R_{\mathrm{ms}}}\big)^k$, therefore we assume $C^{(k>3)}_{\mathrm{b.c}}\rightarrow 0$. It may be convenient to introduce the angle dependent functions $a_k(\theta_{ij})$ as follows
\begin{equation}\label{fKalist}
	\begin{aligned}
		a_3(\theta_{ij}) &=(1/8)(-45\cos^4{\theta_{ij}}+36\cos^2{\theta_{ij}}-3), \\
		a_6(\theta_{ij}) &=(1/4)(-36\cos^4{\theta_{ij}}+9\cos^2{\theta_{ij}}+3),  \\
		a_8(\theta_{ij}) &= (3/4)(-30\cos^4{\theta_{ij}}+2\cos^2{\theta_{ij}}+4), \\
		a_9(\theta_{ij}) &=(9/8)(-33\cos^4{\theta_{ij}}+18\cos^2{\theta_{ij}}-1).
	\end{aligned}
\end{equation}
Those functions, Eq.~(\ref{fKalist}), can be used to rewrite the magnetostriction factor as a series containing contributions from all coordination spheres $ns$ inside the micro-sphere
\begin{equation}\label{MagnFactCompact}
\!\!\!\!	K^{\mathrm{affine}}_{\mathrm{micro}}(k)|_{\mathrm{distr.}}=\frac{4b_k}{3\phi}\!\!\sum_{r^{(0)}_{\mathrm{ns}}<R_{\mathrm{ms}}}\!\!\!\! N_{\mathrm{ns}}\Big(\frac{a}{r^{(0)}_{\mathrm{ns}}}\Big)^k\langle a_k(\theta_{ij}) \rangle_{\mathrm{ns}}^{\mathrm{distr.}},
\end{equation}
where the constants $b_k$ are equal to $1,\beta_1, \beta_2$ and $\beta^2_1$ for $k=3,6,8$ and $k=9$, respectively. The averaging in $\langle a_k(\theta_{ij})\rangle^{\mathrm{distr.}}_{\mathrm{ns}}=\frac{1}{N_{\mathrm{ns}}}\sum_{r^{(0)}_{ij}\in r_{\mathrm{ns}}} a_k(\theta_{ij})$, see Eq.~(\ref{MagnFactCompact}), is performed over the $n$-th coordination sphere. Since those average values are dependent on the distribution pattern of the particles inside an elastomer, so does the magnetostriction factor $K^{\mathrm{affine}}_{\mathrm{micro}}(k)|_{\mathrm{distr.}}$. In order to gain some qualitative insight, it may be instructive to investigate the angular dependency of the magnetostriction factor for the first coordination sphere $r^{(0)}_{\mathrm{1s}}$. One can note that the volume fraction of the particles $\phi\propto (\frac{a}{r^{(0)}_{\mathrm{1s}}})^3$, thus, the rescaled functions, similar to those in Eq.~(\ref{MAngularDeffMain}), can be introduced
\begin{equation}\label{mtheta9}
	m(r^{(0)}_{\mathrm{1s}}; \theta)=\sum_{k=3}^9m_k(r^{(0)}_{\mathrm{1s}}; \theta),
\end{equation}
where
\begin{equation}
	m_k(r^{(0)}_{\mathrm{1s}}; \theta)= b_k\big(\frac{a}{r^{(0)}_{\mathrm{1s}}}\big)^{k-3} a_k(\theta).
\end{equation}
In the Eq.~(\ref{mtheta9}) we sum the first four nonzero contributions ($k=3,6,8$ and $9$ see Eq.~(\ref{fKalist})).
\begin{figure}[h!]
\centering
  \includegraphics[scale=0.69]{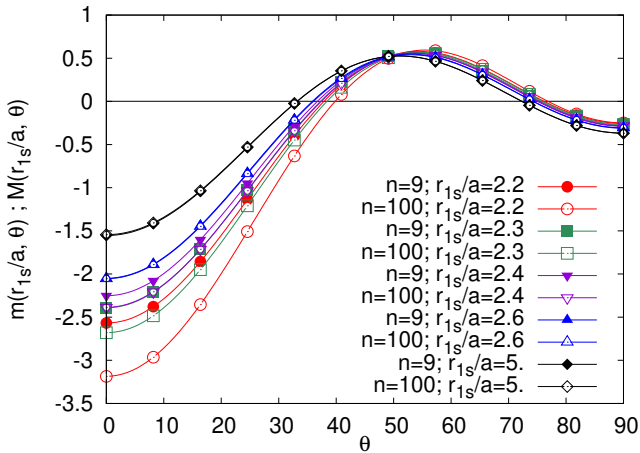}
\caption{Comparison of the angular dependencies for the functions  $m(r^{(0)}_{\mathrm{1s}}; \theta)$ from Eq.~(\ref{mtheta9}) and $M(r^{(0)}_{\mathrm{1s}}; \theta)$ from Eq.~(\ref{MAngularDeffMain}). The former function is derived from the series expansion Eq.~(\ref{LES}) truncated at $n=9$, the latter one originates from the numerical approximation~\cite{Biller2014a, Biller2014} of the series expansion truncated at $n=100$. Several cases for different radii of the first coordination sphere, $r^{(0)}_{\mathrm{1s}}$, are shown.}
 \label{fig:SeriesAng}
\end{figure}
The dependence of $m(r^{(0)}_{\mathrm{1s}}; \theta)$ on the angle $\theta$ is compared to the one considered earlier in Eq.~(\ref{MAngularDeffMain}). The results of the comparison are plotted in the Fig.~(\ref{fig:SeriesAng}). It can be seen that, at the distance between the nearest neighbors $r^{(0)}_{\mathrm{1s}}=5a$, both functional forms are undistinguishable from the prediction of the dipolar potential (see Fig.~(\ref{fig:MAngularMain})). When the radius of the first coordination sphere approaches $r^{(0)}_{\mathrm{1s}}\sim 2.2a-2.3a$, the differences between $m(r^{(0)}_{\mathrm{1s}}; \theta)$ and $M(r^{(0)}_{\mathrm{1s}}; \theta)$ become apparent. For the larger particle separations, $r^{(0)}_{\mathrm{1s}}>2.4a$, these functions are almost identical. Interestingly, the behavior of $M(r^{(0)}_{\mathrm{1s}}; \theta)$ for $r^{(0)}_{\mathrm{1s}}=2.4a$ and $m(r^{(0)}_{\mathrm{1s}}; \theta)$ for $r^{(0)}_{\mathrm{1s}}=2.3a$ looks very similar, especially for small values of the angle $\theta<30^{\circ}$. This may be related to the form of the approximation formula in Eq.~(\ref{Eq1}), where the terms with $k>3$ quickly converge. Thus, the increment of the radius vector values may effectively correspond to the approximation of the series in Eq.~(\ref{LES}) with smaller number of terms $n$. However, we haven't studied this in more detail.

\begin{figure}[h!]
\centering
  \includegraphics[scale=0.69]{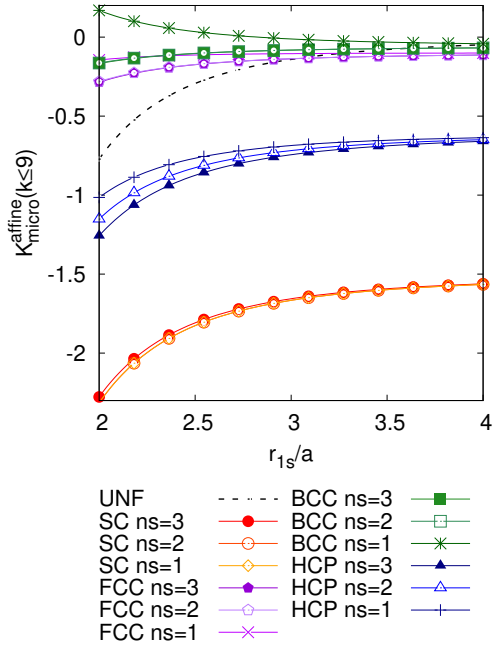}
\caption{Dependency of the function $K^{\mathrm{affine}}_{\mathrm{micro}}(k\leq 9)$ from Eq.~(\ref{Series9K}) on the radius of the first coordination sphere $r^{(0)}_{\mathrm{1s}}$. With the exception of the BCC lattice, at $ns=1$, magnetostriction factor is negative. Additionally, the cases of the uniform (UNF) particles arrangement as well as the SC, FCC and HCP lattices are shown. With the increase of the interparticle separation, $r^{(0)}_{\mathrm{1s}}$, the function quickly converge to the dipole-dipole model predictions.}
 \label{fig:SeriesDistr}
\end{figure}

Now, turning our attention to the Eq.~(\ref{MagnFactCompact}), we can sum up the contribution from the dipolar term, $k=3$, and the known higher order terms, $k=6-9$, to obtain the complete micro-sphere magnetostriction factor
\begin{equation}\label{Series9K}
K^{\mathrm{affine}}_{\mathrm{micro}}(k\leq 9)=K^{\mathrm{affine}}_{\mathrm{micro}}(3)+\sum_{k=6}^9K^{\mathrm{affine}}_{\mathrm{micro}}(k),
\end{equation}
where the first four non-zero terms in the series expansion Eq.~(\ref{LES}) were used~\cite{Jeff}. In a such framework, the inhomogeneous magnetization effects are accounted for in the $k>3$ part of the sum. Since those terms decay with the increase of the distance $r^{(0)}_{\mathrm{ns}}$, we can restrict ourself to the first few coordination spheres only. Contrary to that, the dipolar part of the magnetostriction factor,  $K^{\mathrm{affine}}_{\mathrm{micro}}(3)|_{\mathrm{distr.}}$, exhibits complex behavior, which requires accounting for large number of coordination spheres $r^{(0)}_{\mathrm{ns}}$. This property of the dipolar magnetostriction factor may originate from the interplay of the radial and angular dependent parts, see Eq.~(\ref{MagnFactCompact}). The radial part is proportional to $\propto 1/r^3$, which makes dipolar interactions long-range. However, in the case of uniform distribution, the angular dependent part turns to zero, which makes each term in the sum from Eq.~(\ref{MagnFactCompact}) equal to zero as well. The dipolar magnetostriction factor was calculated for the simple cubic (SC), body-centered cubic (BCC) and hexagonal close-packed (HCP) lattices inside a micro-sphere $R_{\mathrm{ms}}\gg r^{(0)}_{\mathrm{1s}}$ in Ref.~\cite{Dirk19}. For the face-centered cubic (FCC) lattice we performed respective calculations within the interior of the micro-sphere, given by its radius, $R_{\mathrm{ms}}=100 r^{(0)}_{\mathrm{1s}}$.
The function $K^{\mathrm{affine}}_{\mathrm{micro}}(k\leq 9)$, given by the Eq.~(\ref{Series9K}), is plotted in the Fig.~(\ref{fig:SeriesDistr}).
It can be seen that $K^{\mathrm{affine}}_{\mathrm{micro}}(k\leq 9)$ quickly converges to the respective dipole-dipole model prediction, as the radius of the first coordination sphere increases. The case of UNF distribution was calculated by replacing the summation inside a micro-sphere with the integration, under the same assumptions as in Eq.~(\ref{KaffinemicroThreeIsotrAverageMain}). It can be seen, that for sufficiently large interparticle separation, $r^{(0)}_{\mathrm{1s}}/a>3$, the magnetostriction factor approaches zero, as expected. For larger densities of the particles it became negative. The BCC and FCC lattice distributions are the most sensitive to the number of coordination spheres $ns$ accounted for in the calculation. If only the first coordination sphere is considered, the magnetostriction factor $K^{\mathrm{affine}}_{\mathrm{micro}}(k\leq 9)$ for BCC lattice changes sign. It is positive for the denser systems, $r^{(0)}_{\mathrm{1s}}/a<2.5$, and negative for less dense ones, $r^{(0)}_{\mathrm{1s}}/a>3$. The BCC lattice arrangement leads to the negative magnetostriction factor for all densities, if first two or three coordination spheres are considered. Comparing this with the results for the BCC lattice in Fig.~(\ref{fig:DistributionsMain}), one can see that accounting for more terms ($n=100$) in the series in Eq.~(\ref{LES}) increases the weight of the contribution from the first coordination sphere. Contrary to that, the magnetostriction factor for the SC lattice is almost independent on the number of coordination spheres considered. Some magnetostriction factors depicted in Fig.~(\ref{fig:SeriesDistr}) are smaller than the contribution from the initial shape discussed in Sec.~\ref{III}. For example, for the initially spherical sample one predicts $K^{\mathrm{affine}}_{\mathrm{macro}}=2/5$. As a result, for the FCC lattice distribution with smaller densities, $r^{(0)}_{\mathrm{1s}}/a>2.4$, the ``macro'' term is dominant, predicting the net elongation of the sample, but for higher densities, $r^{(0)}_{\mathrm{1s}}/a<2.3$, the ``micro'' contribution is larger, predicting contraction of an MSE. In the case of the BCC lattice arrangement of particles inside a spherical sample, the elastomer is expected to elongate for all densities. Both the SC and HCP cases lead to the contraction of the spherical sample, upon application of the external magnetic field $\bm{H}_0$. The elastomers with the other initial shapes are considered in the next section, see Fig~(\ref{fig:MagnetostrShape}).

\subsection{Contribution from the dipolar interactions at various volume fractions of magnetic particles}

\begin{figure}[h!]
	\centering
	\includegraphics[width=.9\linewidth]{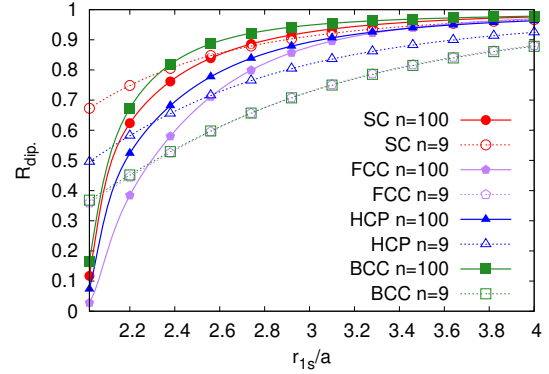}
	\caption{Ratio between the dipolar contribution and the complete micro-sphere magnetostriction factor $R_{\mathrm{dip.}}$, given by the Eq.~(\ref{DipRatio}), as a function of the first coordination sphere radius $r^{(0)}_{\mathrm{1s}}$. Respective ratios $R_{\mathrm{dip.}}$ for various lattices are shown. For each lattice types, two cases are shown. The first one shows $R_{\mathrm{dip.}}$ derived numerically according to Eq.~(\ref{LES}) with $n=100$ terms used (solid lines and symbols). The second one shows the same ratio derived from Eq.~(\ref{LES}) where $n=9$ terms are used explicitly.}\label{fig:Comparison}
\end{figure}

\begin{figure}[h!]
	\centering
	\includegraphics[width=.98\linewidth]{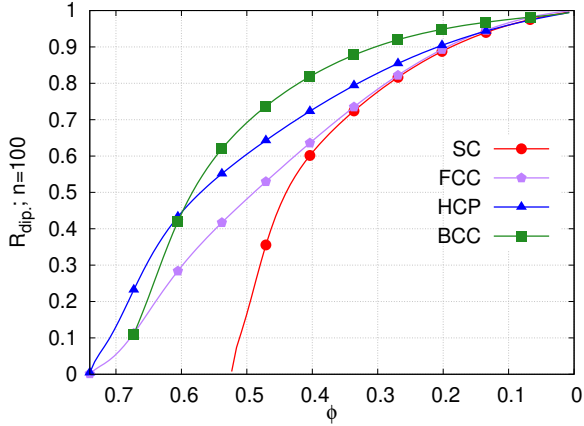}
	\caption{Dependency of the dipolar ratio $R_{\mathrm{dip.}}$, given by the Eq.~(\ref{DipRatio}), on the volume fraction of the particles $\phi$ inside an elastomer is shown. For each lattice, respective plots terminate where the maximal possible packing of spherical particles is reached. In the case of the BCC lattice, the dipolar ratio is always greater than $\approx10\%$.}\label{fig:VolFract}
\end{figure}

Now, let us estimate the influence of inhomogeneous magnetization on the magnetostriction effect. To this end, we introduce the ratio between the dipolar contribution and the complete micro-sphere magnetostriction factor
\begin{equation}\label{DipRatio}
R_{\mathrm{dip.}}=\frac{|K^{\mathrm{affine}}_{\mathrm{micro}}(3)|}{|K^{\mathrm{affine}}_{\mathrm{micro}}(3)|+|K^{\mathrm{affine}}_{\mathrm{micro}}(k>3)|},
\end{equation}
that hereafter will be termed as the ``dipolar ratio''.
When both dipolar and higher order terms in $K^{\mathrm{affine}}_{\mathrm{micro}}$ are of the same sign, it can be rewritten as $K^{\mathrm{affine}}_{\mathrm{micro}}(3)/K^{\mathrm{affine}}_{\mathrm{micro}}(k\geq3)$. In general, dipolar ratio $R_{\mathrm{dip.}}$ is the function of the first coordination sphere radius $r^{(0)}_{\mathrm{1s}}$, and we plot this dependency in the Fig.~(\ref{fig:Comparison}). Two cases are considered: (i) when the dipolar ratio is derived from the numerical approximation formula of the first $n=100$ terms in the series expansion, given by the Eq.~(\ref{LES}); (ii) when it is derived from the first nine ($n=9$) terms explicitly. In both cases, the first three coordination spheres are used in the calculation of the magnetostriction factor higher than the dipolar one,  $K^{\mathrm{affine}}_{\mathrm{micro}}(k>3)$. We should note here that the magnetostriction factor derived from the nine ($n=9$) terms is more sensitive to the number of coordination spheres $ns$ accounted for in calculation, which is also evident comparing Figs.~(\ref{fig:DistributionsMain}) and~(\ref{fig:SeriesDistr}). As before, we use dipolar magnetostriction factor calculated for SC, BCC, HCP lattices in Ref.~\cite{Dirk19}, while FCC lattice result we obtain by summation inside the micro-sphere with the radius $R_{\mathrm{ms}}=100r^{(0)}_{\mathrm{1s}}$. It can be seen that, in $n=100$ case, the inhomogeneous magnetization effects become more pronounced for small values of $r^{(0)}_{\mathrm{1s}}$. Interestingly, the more terms $n=100$ in the Eq.~(\ref{LES}) lead to ``more short-range'' inhomogeneous magnetization effects, than for the series truncated at $n=9$. For the case $n=100$, at distances between the nearest neighbors $r^{(0)}_{\mathrm{1s}}>3a$, the dipolar term comprises more than $90\%$ of magnetostriction factor. This distances correspond to different volume fractions $\phi$ for different lattices, as can be seen in the Table~(\ref{tab:table5}). The dipole ratio $R_\mathrm{dip.}$ dependency on $\phi$ is shown in the Fig.~(\ref{fig:VolFract}) for the $n=100$ case.

It can be seen, in Fig.~(\ref{fig:VolFract}), that for the SC, FCC, and HCP lattices, the dipolar ratio $R_{\mathrm{dip.}}$ is negligible for dense systems. And for the BCC lattice it is always greater than one-tenth of the total magnetostriction factor. The maximal possible value of the volume fraction $\phi_{\mathrm{max}}$ can be reached when the interparticle separation is minimal $r^{(0)}_{\mathrm{1s}}=2a$. In the case of the SC and BCC lattices, the $\phi_{\mathrm{max}}\approx0.52$ and $\approx0.68$, respectively, while for both the HCP and FCC lattices $\phi_{\mathrm{max}}\approx0.74$. In the case of BCC lattice, the dipolar contribution comprises more than $90\%$ of the complete micro-sphere magnetostriction factor for the volume fraction less than $\phi<0.3$. Similar value of dipolar ratio $R_{\mathrm{dip.}}$ is reached when $\phi<0.2$ in the case of the SC, FCC and HCP lattices. The SC and FCC arrangements predict similar dipolar contribution to the magnetostriction effect on the interval $\phi\in(0;0.35)$.

In the current section we showed an influence of the inhomogeneous magnetization on the magnetostriction in the framework of PIA. The effect of finite size of the particles become important when they are sufficiently close to each other. For example, in Fig.~(\ref{fig:Comparison}) it can be seen that, the dipolar contribution comprises more than $90\%$ of the complete magnetostriction factor, when $r^{(0)}_{\mathrm{1s}} > 3.2a$. This interparticle separation corresponds roughly to the volume fraction $\phi<0.2$ for most of the lattices considered, see Fig.~(\ref{fig:VolFract}). The exception being the BCC lattice, where the limiting interparticle separation is $r^{(0)}_{\mathrm{1s}} > 2.8a$ and the respective volume fraction $\phi<0.3$, see Figs.~(\ref{fig:Comparison},\ref{fig:VolFract}). This results are obtained in the framework involving approximation formula for $n=100$ terms of the Laplace equation solution, given by Eq.~(\ref{LES}). They are also compared with the results originated from the explicit treatment of the first $n=9$ terms in Eq.~(\ref{LES}). It can be seen in Fig.~(\ref{fig:Comparison}), that the latter approach underestimates dipolar contribution to magnetostriction effect for larger values of $r^{(0)}_{\mathrm{1s}}$  (less dense distributions). It also overestimates the importance of dipole-dipole interaction when particles become close to each other $r^{(0)}_{\mathrm{1s}}<2.4a$. This may be related to the slow convergence of the sum in Eq.~(\ref{LES}), reported in Refs.~\cite{Biller2014a, Biller2014} earlier. In our analysis we ignore the three- and higher particle interactions and non-additive nature of inhomogeneous magnetization effects, reported for similar electrically polarizable particles~\cite{Bossis93}. However, we speculate that the influence on these interactions with more than two neighboring particles may actually make the magnetization more uniform. And the range of applicability of the dipolar model may increase to the higher volume fractions than those roughly estimated here.

\section{Estimation of the equilibrium strain}\label{V}

In this chapter we would like to estimate the equilibrium strain, given be the Eq.~(\ref{StrainOne}), for the spheroidal MSE with the magnetic particles distributed in the vertices of four different lattices. Similar studies where performed earlier, where only dipolar interactions where accounted for~\cite{Dirk19}. Due to inhomogeneous magnetization of the particles, the dependency of $K^{\mathrm{affine}}_{\mathrm{micro}}$ on the volume fraction $\phi$ may arise. This is expected to create richer magnetostriction behavior of the MSEs. In our framework we accounted for inhomogeneous magnetization in PIA, which may not be accurate enough in order to make confident predictions about behavior of the MSE. The aim is, however, to show the connection between our theoretical model and measurable macroscopic properties of the sample. And also to demonstrate that the equilibrium strain dependency on the volume fraction, $\varepsilon_{eq}\propto \phi^2$, may deviate from the quadratic law in principle, since it can be interpreted as a manifestation of the inhomogeneous magnetization of the particles.

\begin{figure*}
	\centering
	
	\includegraphics[width=.4\linewidth]{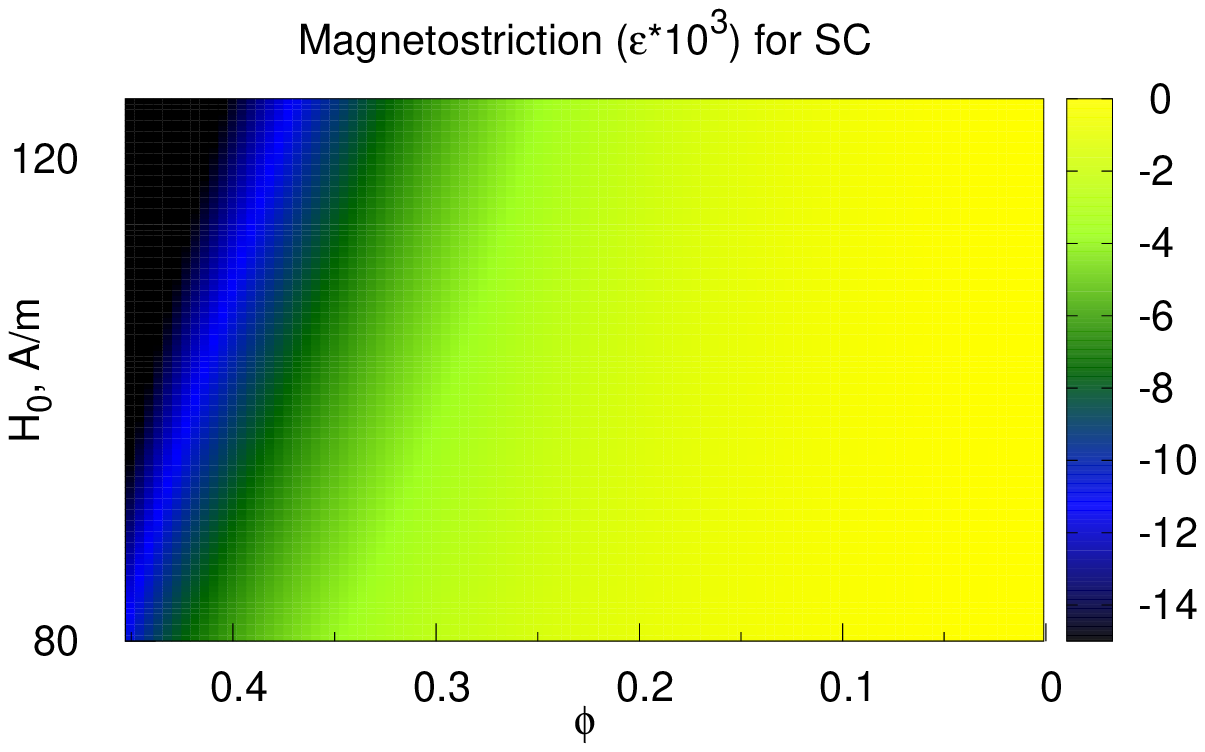}
	\includegraphics[width=.4\linewidth]{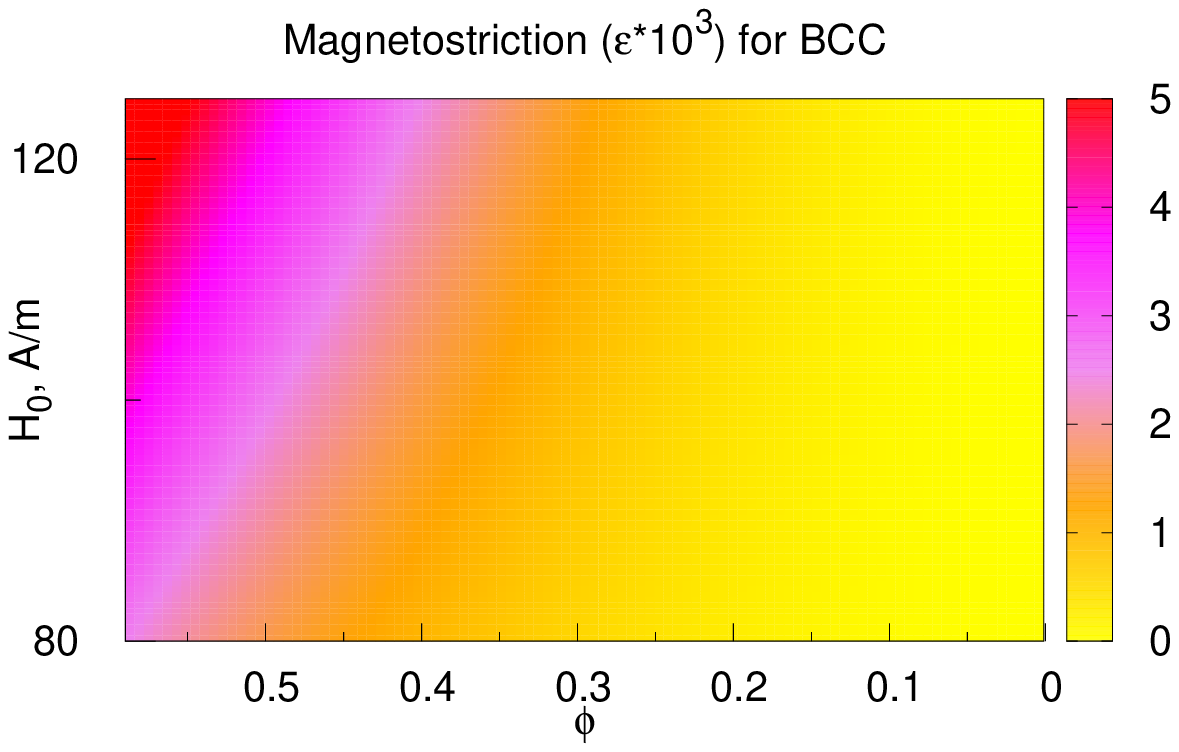}
	\medskip
	\includegraphics[width=.4\linewidth]{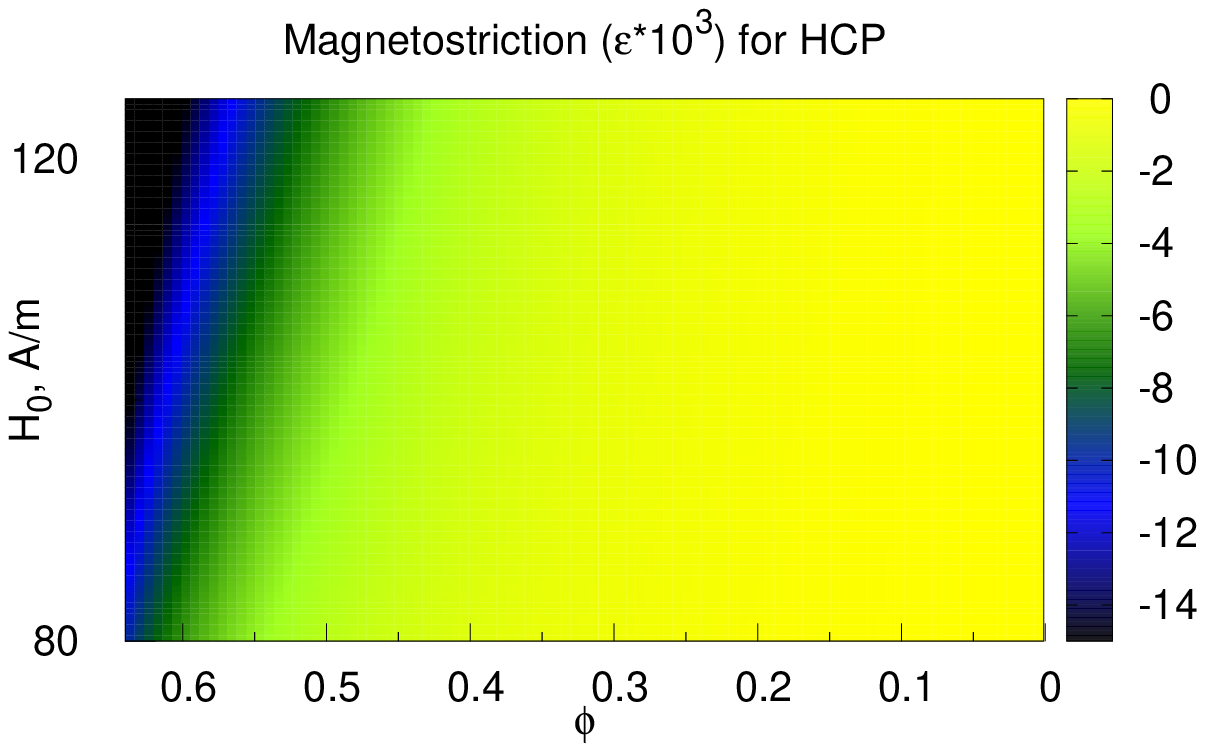}
	\includegraphics[width=.4\linewidth]{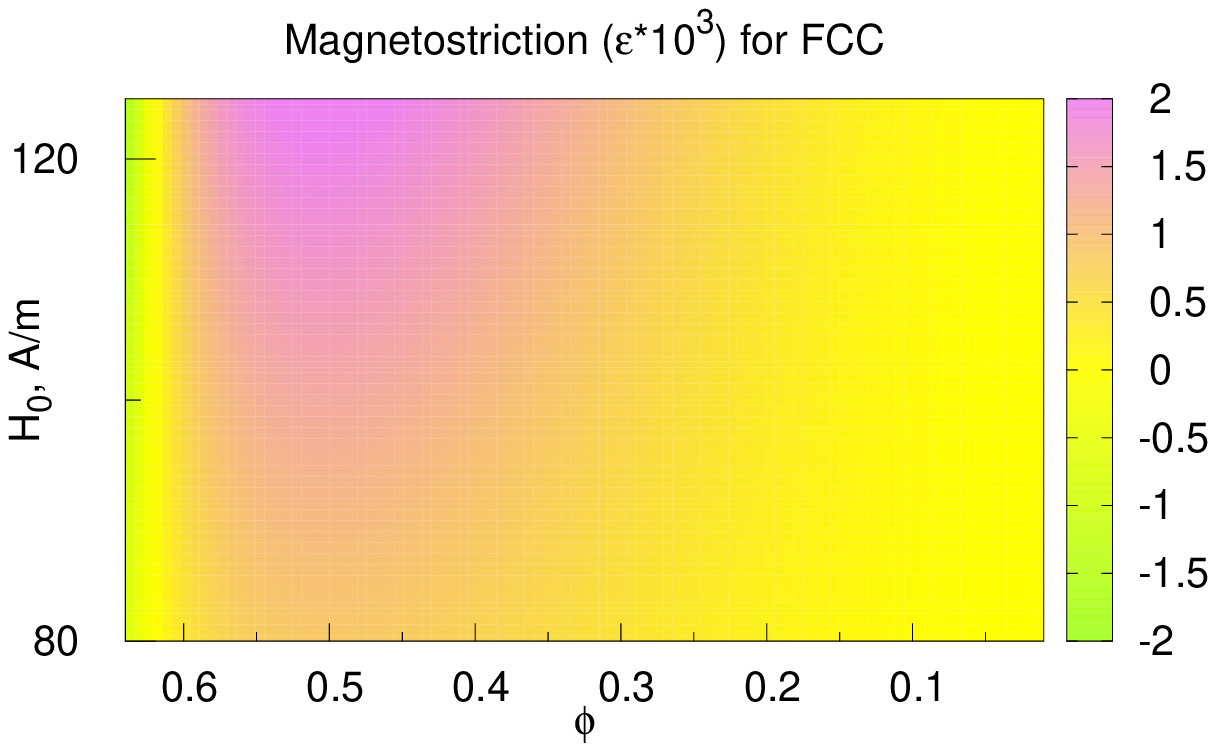}
	\caption{The values of the equilibrium strain, given by the Eq.~(\ref{StrainOne}), for varying strength of the magnetic field $H_0$ and volume fraction of particles $\phi$. The initial aspect ratio $\gamma_0=1$ and the Young's modulus $E=2$ Mpa of the elastomer are fixed.}
	
	\label{fig:MagnetostrField}
\end{figure*}

To this end, we use the values of the magnetostriction factors $K^{\mathrm{affine}}_{\mathrm{micro}}$ calculated in the previous section. Additionally, we specify some physical parameters. We would consider the sample with fixed effective Young's modulus $E=2$ MPa. The magnetic field applied along the axis of the spheroidal sample is considered in the range $H_0\in [80;125]$ kA/m, which is below the saturation field for carbonyl iron particles. The volume fraction of the particles $\phi\propto (a/r^{(0)}_{\mathrm{1s}})^3$ is another varying parameter. Also we consider different aspect ratios of the initial shape of the spheroid $\gamma_0\in[0.1;4]$. The quantitative measure of the magnetostriction effect is the equilibrium strain $\varepsilon_{\mathrm{eq}}$, which the elastomer attains when the external magnetic field is applied.

\begin{figure*}
	\centering
	
	\includegraphics[width=.4\linewidth]{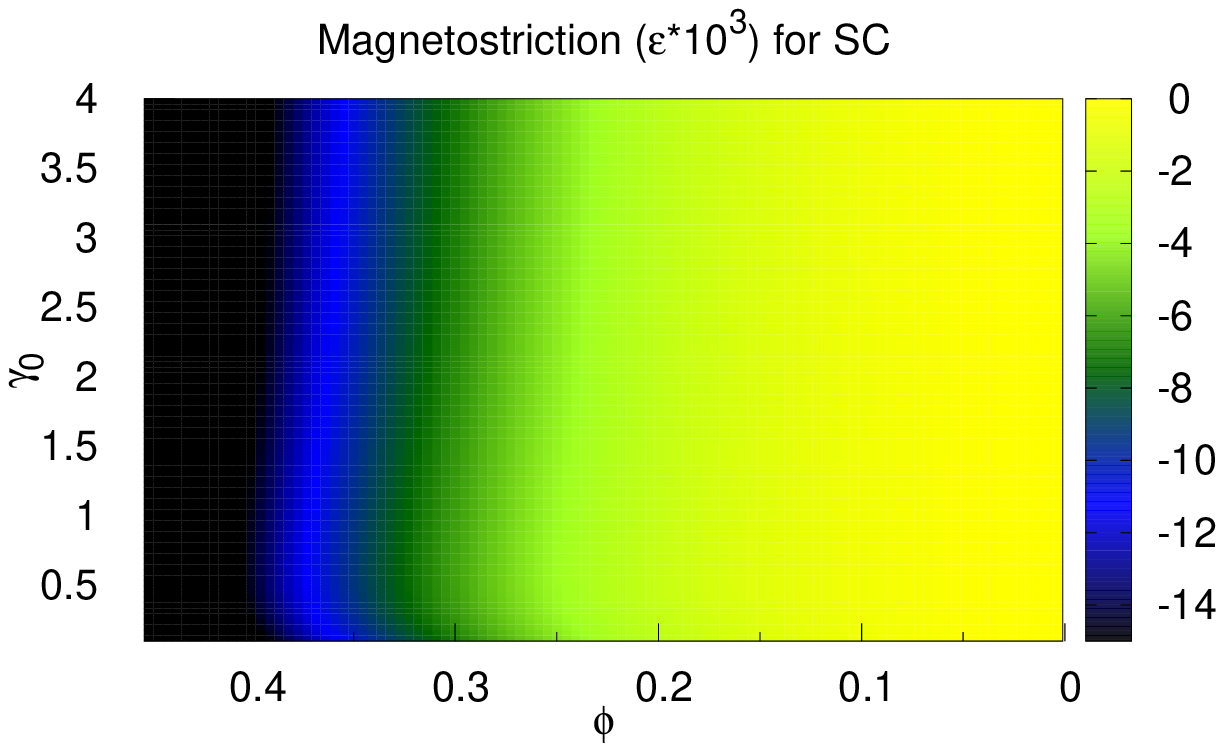}
	\includegraphics[width=.4\linewidth]{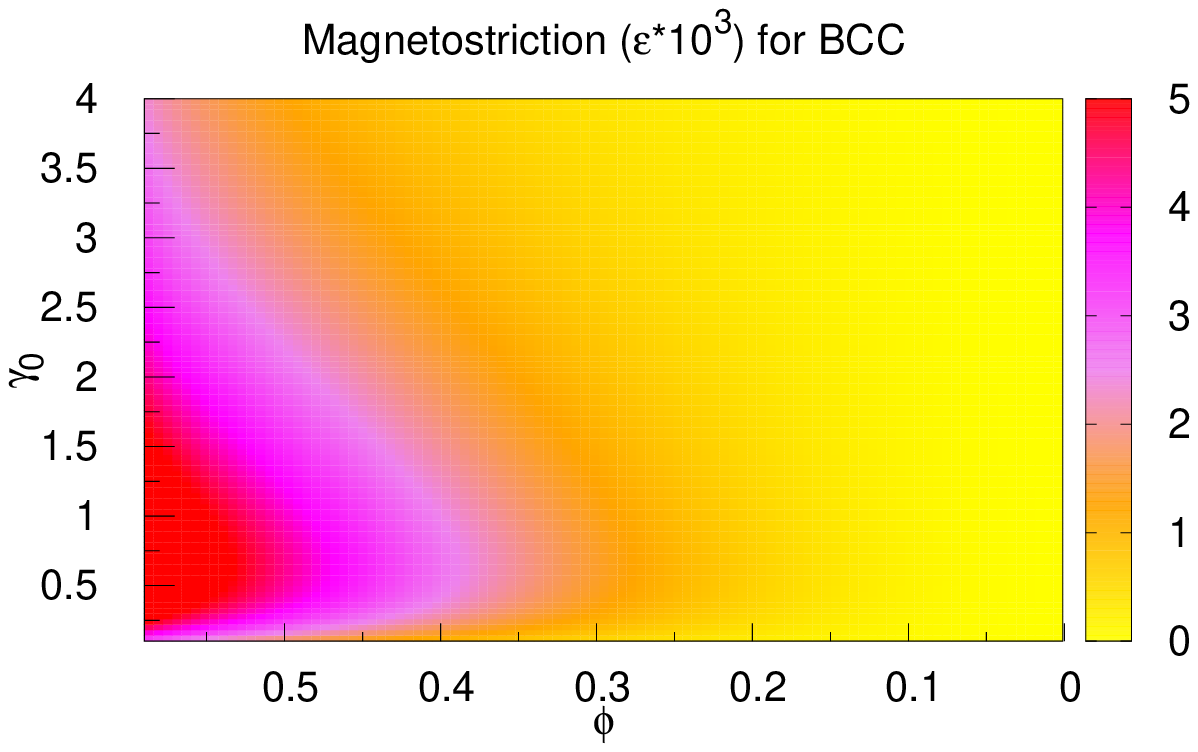}
	\medskip
	\includegraphics[width=.4\linewidth]{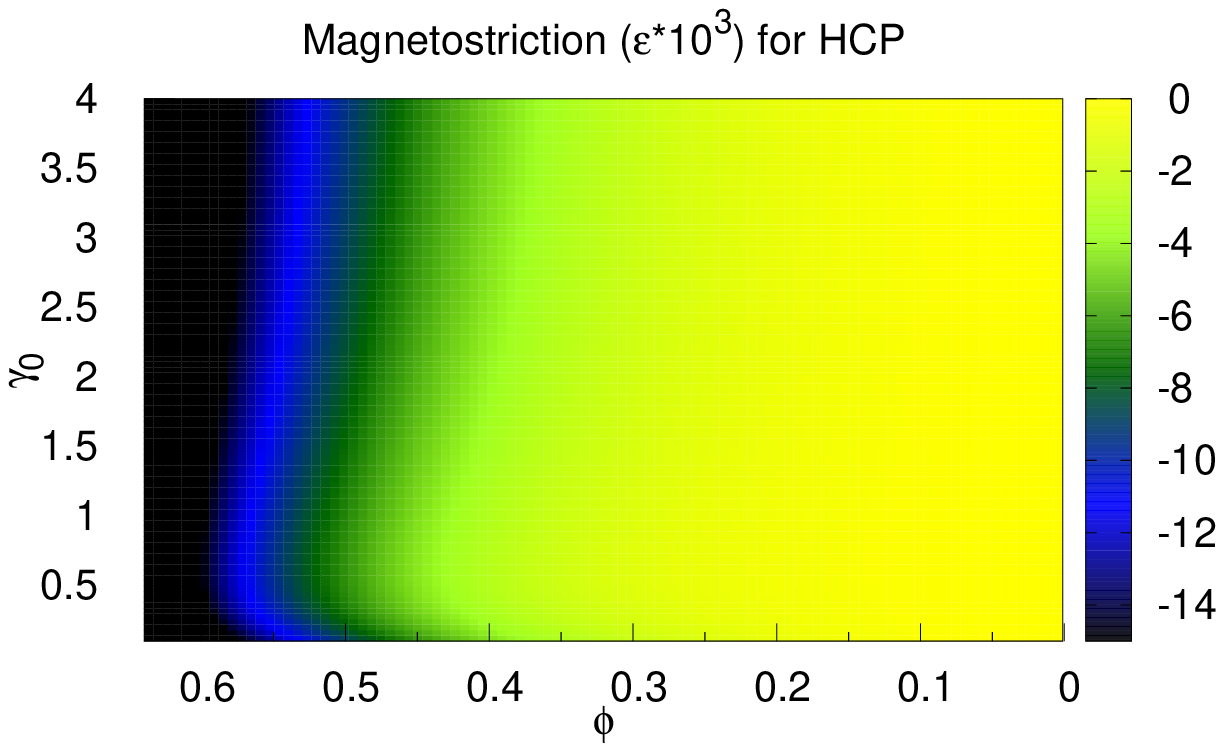}
	\includegraphics[width=.4\linewidth]{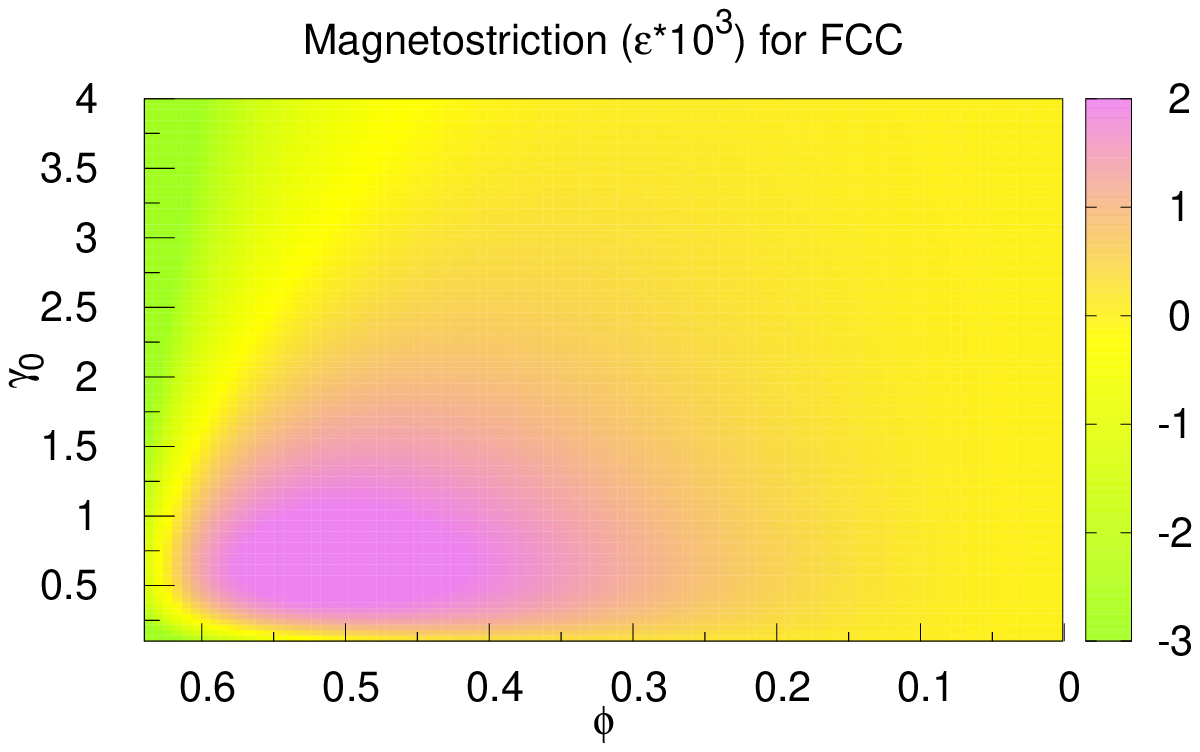}
	\caption{The values of the equilibrium strain, given by the Eq.~(\ref{StrainOne}), for varying initial aspect ratio $\gamma_0$ and the volume fraction $\phi$. The magnetic field strengths and Young's modulus of the elastomer are fixed at $H_0=125$~kA/m and $E=2$~MPa, respectively.}
	
	\label{fig:MagnetostrShape}
\end{figure*}

The dependencies of the equilibrium strain, $\varepsilon_{\mathrm{eq}}$, on the volume fraction of particles $\phi$ and on the magnitude of the external magnetic field $H_0$ are shown in Fig.~(\ref{fig:MagnetostrField}). It can be seen, that the magnitude of the effect is lower than several percents. For higher values of the field and larger volume fractions of the particles, the equilibrium strain for the SC and HCP lattices reaches $2\%$. The sign of the magnetostriction effect for those two lattices is negative. For the BCC lattice, one can see that the equilibrium strain is of smaller amplitude, reaching $0.5\%$, while the sign is positive. In the case of the FCC lattice, the magnitude of the equilibrium strain is $0.2\%$, and the sign of the magnetostriction effect is different for different set of the parameters. For the intermediate volume fractions the sign is positive, while for very densely packed particles the effect is predicted to be negative. We want to note, that FCC lattice is considered positioned in a way that the magnetic field is directed parallel to the side of the cube. If we considered FCC lattice rotated to the close-packed conformation, where direction of the field would be collinear to the diagonal of the cube, the predicted magnetostriction effect would be more similar to the one shown for the HCP lattice case.

The dependency of the equilibrium strain, $\varepsilon_{\mathrm{eq}}$, on the initial aspect ratio $\gamma_0$ of a sample and the volume fraction $\phi$ for each lattice is shown in the Fig.~(\ref{fig:MagnetostrShape}). The effective Young's modulus of a sample is assumed to be fixed, and equal to $E=2$ MPa. It can be seen that the SC and HCP lattices are similar and predict negative magnetostriction effect with $\varepsilon_{\mathrm{eq}}$ larger than $2\%$ and $1\%$, respectively. The equilibrium strain decreases significantly as the volume fraction of the particles, $\phi$, becomes smaller. For the BCC lattice, our model predicts a positive sign for the magnetostriction, and the effect is more pronounced for the slightly oblate spheroids. In the case of the FCC lattice, the oblate spheroids would elongate along the applied magnetic field, positive magnetostriction. But the prolate initial shape of the elastomer would lead to the contraction of the sample along applied field, negative magnetostriction. The magnitude of the effect is again smaller than for the other lattices reaching $\varepsilon_{\mathrm{eq}}\sim 0.2\%$.

\section{Conclusions}

In the current paper we start from the magnetic energy of the pair of linearly magnetized spheres. At small interparticle separations these particles inhomogeneously magnetize each other. In order to account for such inhomogeneous magnetization one need to solve the Laplace equation for this spherical particles. The approximate formula, designed to fit the solution of the Laplace equation, given by the Eq.~(\ref{LES}) and truncated at its hundredth term, was reported in~\cite{Biller2014}. We use this formula in the PIA to construct the magnetic energy density of the spheroidal elastomer. Then, we study the effect of the magnetostriction for the MSE sample upon application  of the external magnetic field $\bm{H}_0$. The direction of $\bm{H}_0$ is chosen to be collinear to the symmetry axis of the sample.
The elastic energy density of the elastomer is obtained assuming that the sample behaves as a spring with Young's modulus $E$. In our model, the magnetic energy density, as well as the interparticle distances, depend on the strain $\varepsilon$ of the elastomer. The equilibrium strain $\varepsilon_{\mathrm{eq}}$ is evaluated within the linear approximation for the magnetic energy density.
Then, we analyze how the initial shape and the distribution of particles inside the MSE affect sign and amplitude of the equilibrium strain.

The contribution to the magnetostriction effect from the initial shape of the sample is determined by the magnetostriction factor $K_{\mathrm{macro}}$, (see Eq.~(\ref{KmicroKmacro})). In the current study we show that only the dipolar interactions provides a significant contribution to the $K_{\mathrm{macro}}$, as long as MSE is much larger than the particle radius. This result is in agreement with the continuum mechanics approach.

The magnetostriction factor inside a micro-sphere, $K^\mathrm{affine}_\mathrm{micro}$, describes the contribution to the magnetostriction effect from the initial distribution of the particles. We found that the interactions of a higher order than dipolar, originated from the effects of inhomogeneous magnetization, are short ranged. Thus, for volume fractions of the particles, $\phi<0.2$, in the case of SC, FCC, and HCP lattices, the dipolar term comprises more than $90\%$ of the magnetostriction factor. The same is true for $\phi<0.3$, in the case of BCC lattice, as can be seen Fig.~(\ref{fig:VolFract}).

At small interparticle separations, when first coordination sphere radius is less than $r^{(0)}_{1s}\approx3a$, the inhomogeneous magnetiation of the particles leads to $K^\mathrm{affine}_\mathrm{micro}$ dependence on the volume fraction of the particles, $\phi$. Since such behavior is not predicted when only dipolar interactions are taken into account, it may be thought as the manifestation of the effect of inhomogeneous magnetization. Consequences of this can be observed in Figs.~(\ref{fig:MagnetostrField},\ref{fig:MagnetostrShape}), where dependence of the equilibrium strain, $\varepsilon_\mathrm{eq}$, on the volume fraction, $\phi$, is shown.

Since $K^\mathrm{affine}_\mathrm{micro}$ contains terms proportional to $\cos^4{\theta_{ij}}$, it is dependent on the orientation of the lattice relatively to the direction of the magnetic field $\bm{H}_0$. Such effect can be illustrated by the differences between the FCC and HCP lattices, Figs.~(\ref{fig:MagnetostrField},\ref{fig:MagnetostrShape}). The orientation of FCC lattice was chosen such that the edge of the cube is collinear with $\bm{H}_0$. If the field direction was collinear with its diagonal, the differences between those lattices would have been minor.

When particles arrangement is close to a close contact, $r^{(0)}_{1s}\approx2a$, the SC, FCC and HCP lattices predict giant negative values of magnetostriction factor $K^\mathrm{affine}_\mathrm{micro}$, see Fig.~(\ref{fig:DistributionsContact}). We treat this as an artefact of our model. We assume strong particle-matrix coupling and affine deformation of the sample. At short distances, when a small amount of elastic component is present, this assumption is not valid any more. Additionally, our magnetic energy density is derived under the pairwise interaction approximation, which in the case of inhomogeneously magnetized particles, may not be sufficiently accurate~\cite{Bossis93}.

In order to make our model more realistic, it is important to consider other laws of elastic reaction to the microscopic movement of magnetic particles. Also, energy of multiple inhomogeneously magnetized particles should be considered without pairwise interaction approximation. This may shed light onto the limitations and the range of applicability of our model. Also, it would be interesting to calculate equilibrium strain above linear approximation of magnetic energy. This would led to additional effects, for example, for the change of an effective Young modulus as magnetic particles rearrange under application of the external field.

\noindent\textbf{Acknowledgements}\\
D.Y. thanks the National Academy of Sciences of Ukraine for funding this research (the grant for research laboratories/groups of young scientists No 07/01-2022(4)).\\
D.Y. and J.I are grateful to the Armed Forces of Ukraine for the protection during this research work.

\bibliographystyle{is-unsrt}
%\bibliography{NeoHooke}
%\bibliographystyle{unsrt}
\bibliography{PaperR}

%\newpage

\appendix

\onecolumn

%%% For Appendix A.
% format the equation environment
\renewcommand{\theequation}{A.\arabic{equation}}

% reset the counter
\setcounter{equation}{0}

% appendix section. * is used to suppress the section numbering
%\section*{Appendix B. <some heading here>}
\section*{Appendix A. Integral}
We are interested in calculating the integral
\begin{equation}\label{AppIntegral}
F_{\mathrm{macro}}^{(k>3)}=\beta_1^{p_k-2}\int_0^1 dx \int_{R_{\mathrm{ms}}}^{r_x}dr r^2\bigg[     \frac{a^{k-3}A_k}{(r-aB_k)^k}+\frac{a^{k-3}C_k}{(r-aD_k)^k}x^2   \bigg].
\end{equation}
Let us consider the following integral first
\begin{equation*}
\int_{R_{\mathrm{ms}}}^{r_x} drr^2 \frac{a^{k-3}A_k}{(r-aB_k)^k}=a^{k-3}A_k\int_{R_{\mathrm{ms}}-aB_k}^{r_x-aB_k}dy\frac{(y+aB_k)^2}{y^k}=
\end{equation*}
\begin{equation*}
=\frac{A_k}{3-k}\Big[\frac{a^{k-3}}{(r_x-aB_k)^{k-3}}-\frac{a^{k-3}}{(R_{\mathrm{ms}}-aB_k)^{k-3}}\Big]+\frac{2A_kB_k}{2-k}\Big[\frac{a^{k-2}}{(r_x-aB_k)^{k-2}}-\frac{a^{k-2}}{(R_{\mathrm{ms}}-aB_k)^{k-2}}\Big]+
\end{equation*}
\begin{equation}\label{AppIn}
+\frac{A_kB^2_k}{1-k}\Big[\frac{a^{k-1}}{(r_x-aB_k)^{k-1}}-\frac{a^{k-1}}{(R_{\mathrm{ms}}-aB_k)^{k-1}}\Big].
\end{equation}
Or
\begin{equation}
\int_{R_{\mathrm{ms}}}^{r_x} drr^2 \frac{a^{k-3}A_k}{(r-aB_k)^k}=\sum_{l=1}^3 2^{1-|l-2|}\frac{A_kB_k^{3-l}}{l-k}\Big[\frac{a^{k-l}}{(r_x-aB_k)^{k-l}}-\frac{a^{k-l}}{(R_{\mathrm{ms}}-aB_k)^{k-l}}\Big].
\end{equation}
And the original integral
\begin{equation}\label{AppFtogether}
F_{\mathrm{macro}}^{(k>3)}=\beta_1^{p_k-2}\sum_{l=1}^3 \frac{2^{1-|l-2|}}{l-k}\Big(\int_0^1dx\Big[\frac{a^{k-l}A_kB_k^{3-l}}{(r_x-aB_k)^{k-l}}+\frac{a^{k-l}C_kD_k^{3-l}x^2}{(r_x-aD_k)^{k-l}}\Big]-\frac{a^{k-l}A_kB_k^{3-l}}{(R_{\mathrm{ms}}-aB_k)^{k-l}}-\frac{1}{3}\frac{a^{k-l}C_kD_k^{3-l}}{(R_{\mathrm{ms}}-aD_k)^{k-l}}\Big)
\end{equation}
As long as $\frac{a}{A}\ll1$, the terms which contain $r_x$ can be neglected.
First, one can expand the denominator into Taylor series to get
\begin{equation}\label{AppTaylor}
\int_0^1dx\frac{a^{k-l}}{(r_x-aB_k)^{k-l}}\rightarrow \big(\frac{a}{A}\big)^{k-l} \int_0^1dx \big[(1-\gamma_\varepsilon^2)x^2+\gamma_\varepsilon^2\big]^{(k-l)/2}\bigg(1+\frac{aB_k(k-l)}{A}\big[(1-\gamma_\varepsilon^2)x^2+\gamma_\varepsilon^2\big]^{1/2}+\cdots  \bigg).
\end{equation}
Than, multiplying the factors $\bigg(\frac{a}{A}\bigg)^n$ to the unity in the form $1=\frac{1+\gamma_\varepsilon^n}{1+\gamma_\varepsilon^n}$, one get the sum of the terms like the following
\begin{equation}\label{AppTaylorTerm}
\bigg[\bigg(\frac{a}{A}\bigg)^{n}+\bigg(\frac{a}{B}\bigg)^{n}\bigg]\frac{1}{1+\gamma_\varepsilon^{n}}\int_0^1dx\big[(1-\gamma_\varepsilon^2)x^2+\gamma_\varepsilon^2\big]^{n/2}.
\end{equation}
Similar procedure can be done for $\int_0^1dxx^2\frac{a^{k-l}}{(r_x-aD_k)^{k-l}}$. The corresponding
functions $\frac{1}{1+\gamma_\varepsilon^{n}}\int_0^1dx\big[(1-\gamma_\varepsilon^2)x^2+\gamma_\varepsilon^2\big]^{n/2}$ and $\frac{1}{1+\gamma_\varepsilon^{n}}\int_0^1dxx^2\big[(1-\gamma_\varepsilon^2)x^2+\gamma_\varepsilon^2\big]^{n/2}$, emerging in the process, are bound for all $\gamma_\varepsilon$. Thus, for macroscopic sample condition
\begin{equation}
 \bigg[\bigg(\frac{a}{A}\bigg)^{n}+\bigg(\frac{a}{B}\bigg)^{n}\bigg]\ll 1,
\end{equation}
the following integrals tend to zero
\begin{equation}
\int_0^1dx\frac{a^{k-l}}{(r_x-aB_k)^{k-l}}\rightarrow 0;\qquad \int_0^1dx\frac{a^{k-l}x^2}{(r_x-aD_k)^{k-l}}\rightarrow 0.
\end{equation}
As can be seen from the last term in Eq.~(\ref{AppFtogether}) and the assumption that radius of the microsphere is in order of ten interparticle distances ($R_{\mathrm{ms}}\ge10\langle r_{ij}\rangle$), one can obtain $\frac{a}{R_{\mathrm{ms}}}<0.05$ at least, with $A_4=0$, $C_4=0.0342$ the largest term in $F_{\mathrm{macro}}^{(k>3)}$ would be of the order $10^{-3}$ or smaller. We want to note also that the lower bound of the integral would be the upper bound of the sum, thus $F(\varepsilon)=F_{\mathrm{micro}}(\varepsilon)+F_{\mathrm{macro}}(\varepsilon)$ from Eq.~(\ref{FmFm}) does not actually depend on the $R_{\mathrm{ms}}$ if it is large enough to allow sum interchange with integral.

%%% For Appendix A.
% format the equation environment
\renewcommand{\theequation}{B.\arabic{equation}}

% reset the counter
\setcounter{equation}{0}

% appendix section. * is used to suppress the section numbering
%\section*{Appendix B. <some heading here>}
\section*{Appendix B. Higher $K_{\mathrm{macro}}$}

Higher than ordinary dipole terms in $F_{\mathrm{macro}}^{(k>3)}$ may lead to the higher terms in the factor $K_{\mathrm{macro}}^{(k>3)}=\big[\frac{\partial}{\partial \varepsilon} F_{\mathrm{macro}}^{(k>3)}\big]_{\varepsilon=0}$. It turns out, that for affine deformations this higher terms are zero in the macroscopic sample limit. To check this, one may note first, that terms containing $R_{\mathrm{ms}}$ in the denominator in Eq.~(\ref{AppFtogether}) do not depend on the strain $\varepsilon$, and vanish after taking the derivative. And since $\big[\frac{\gamma_\varepsilon}{\partial \varepsilon}\big]_{\varepsilon=0}=\frac{3}{2}\gamma_0$, it is left to show that
\begin{equation}
\gamma_0\frac{\partial}{\partial \gamma_0}\int_0^1dx\frac{a^{k-l}}{\big(r_x(\gamma_0)-aB_k\big)^{k-l}}\rightarrow 0;\qquad \gamma_0\frac{\partial}{\partial \gamma_0}\int_0^1dx\frac{a^{k-l}x^2}{\big(r_x(\gamma_0)-aD_k\big)^{k-l}}\rightarrow 0.
\end{equation}
Using $\frac{a}{A}\ll 1$ as in Eq.~(\ref{AppTaylor}), one can get Taylor series to obtain
\begin{equation}\label{AppTaylorSum1}
\gamma_0\frac{\partial}{\partial \gamma_0}\int_0^1dx\frac{a^{k-l}}{(r_x(\gamma_0)-aB_k)^{k-l}}\rightarrow \sum_{n=1}C^{B_k}_n\bigg[\bigg(\frac{a}{A}\bigg)^{n}+\bigg(\frac{a}{B}\bigg)^{n}\bigg] \frac{1}{1+\gamma_0^{n}}\gamma_0\frac{\partial}{\partial \gamma_0} \int_0^1dx\big[(1-\gamma_0^2)x^2+\gamma_0^2\big]^{n/2}
\end{equation}
and
\begin{equation}\label{AppTaylorSum2}
\gamma_0\frac{\partial}{\partial \gamma_0}\int_0^1dx\frac{a^{k-l}x^2}{(r_x(\gamma_0)-aD_k)^{k-l}}\rightarrow \sum_{n=1}C^{D_k}_n\bigg[\bigg(\frac{a}{A}\bigg)^{n}+\bigg(\frac{a}{B}\bigg)^{n}\bigg]\frac{1}{1+\gamma_0^{n}}\gamma_0\frac{\partial}{\partial \gamma_0}\int_0^1dxx^2\big[(1-\gamma_0^2)x^2+\gamma_0^2\big]^{n/2},
\end{equation}
where $C^{B_k}_n$ and $C^{D_k}_n$ are constants and the prefacrots $\big(\frac{a}{A}\big)^n$ are multiplied by $\frac{1+\gamma_0^n}{1+\gamma_0^n}$.
The last step is to show that the functions on $\gamma_0$ in the right-hand side of Eqs.~(\ref{AppTaylorSum1},\ref{AppTaylorSum2}) are bounded for all $\gamma_0 \geq 0$.
Let us consider the function from Eq.~(\ref{AppTaylorSum1}) as follows
\begin{equation}
\frac{\gamma_0}{1+\gamma_0^{n}}\frac{\partial}{\partial \gamma_0}\int_0^1dx\big[(1-\gamma_0^2)x^2+\gamma_0^2\big]^{n/2}=\frac{n}{2}\frac{2\gamma_0^2}{1+\gamma_0^n}\int_0^1dx(1-x^2)\big[(1-\gamma_0^2)x^2+\gamma_0^2\big]^{(n-2)/2}.
\end{equation}
This function is bounded for large $\gamma_0$. For small $\gamma_0$ it is bounded for $n\geq 2$. In the case of $n=1$ and small $\gamma_0$ we need to perform integration first, to avoid the dependency on $x$ before taking derivative and taking small $\gamma_0$ limit. The integral would be as follows
\begin{equation}
 \int_0^1dx\big[(1-\gamma_0^2)x^2+\gamma_0^2\big]^{1/2}=\frac{1}{2}+\frac{1}{2}\frac{\gamma_0^2}{\sqrt{1-\gamma_0^2}}\mathrm{Arcth}\big(\sqrt{\gamma_0^{-2}-1}\big).
\end{equation}
The corresponding derivative is
\begin{equation}
\gamma_0\frac{\partial}{\partial \gamma_0} \int_0^1dx\big[(1-\gamma_0^2)x^2+\gamma_0^2\big]^{1/2}=\frac{\gamma_0^2}{2}\frac{2-\gamma_0^2}{(1-\gamma_0^2)^{3/2}}\mathrm{Arcth}\big(\sqrt{\gamma_0^{-2}-1}\big)-\frac{1}{2}\frac{\gamma_0^2}{1-\gamma_0^2}.
\end{equation}
Using L'H\^{o}pital's rule, one get $\lim_{\gamma_0\rightarrow 0}\gamma^2_0\mathrm{Arcth}\big(\sqrt{\gamma_0^{-2}-1}\big)\rightarrow 0$, and thus
\begin{equation}
\lim_{\gamma_0\rightarrow 0}\frac{1}{1+\gamma_0}\gamma_0\frac{\partial}{\partial \gamma_0} \int_0^1dx\big[(1-\gamma_0^2)x^2+\gamma_0^2\big]^{1/2}\rightarrow 0.
\end{equation}
As a result, the function in Eq.~(\ref{AppTaylorSum1}) is bounded for all $\gamma_0\geq0$.
Now, let us consider the function from Eq.~(\ref{AppTaylorSum2}) as follows
\begin{equation}
\frac{\gamma_0}{1+\gamma_0^{n}}\frac{\partial}{\partial \gamma_0}\int_0^1dxx^2\big[(1-\gamma_0^2)x^2+\gamma_0^2\big]^{n/2}=\frac{n}{2}\frac{2\gamma_0^2}{1+\gamma_0^n}\int_0^1dxx^2(1-x^2)\big[(1-\gamma_0^2)x^2+\gamma_0^2\big]^{(n-2)/2}.
\end{equation}
This function is also bounded for all $\gamma_0\geq0$.
Finally, for the macroscopic sample limit $ \bigg[\bigg(\frac{a}{A}\bigg)^{n}+\bigg(\frac{a}{B}\bigg)^{n}\bigg]\ll 1$ one can write
\begin{equation}\label{AppKFinal}
\bigg[\frac{\partial}{\partial \varepsilon} F_{\mathrm{macro}}^{(k>3)}\bigg]_{\varepsilon=0}\rightarrow 0.
\end{equation}
One of the consequences is that the border crossing terms defined in Ref.~\cite{Dirk19} can be calculated from the dipole interactions while others can be ignored. Using the same procedure as in Ref.~\cite{Dirk19}, the border crossing terms explicitly can be written as follows
\begin{equation}\label{AppBorderCrossingDipole}
C_{\mathrm{b.c}}=\frac{3}{2}\bigg[\gamma_0\frac{\partial}{\partial \gamma_0}F_{\mathrm{macro}}(\gamma_0)\bigg]_{\gamma_0=1}=-\frac{3}{2}\int_0^1dx (3x^2-1)(1-x^2)=\frac{2}{5}.
\end{equation}

%%% For Appendix A.
% format the equation environment
\renewcommand{\theequation}{C.\arabic{equation}}

% reset the counter
\setcounter{equation}{0}

% appendix section. * is used to suppress the section numbering
%\section*{Appendix B. <some heading here>}
\section*{Appendix C. Short note on the averaging of the cosine in the lattices}

\begin{figure}[h!]
\centering
  \includegraphics[scale=0.69]{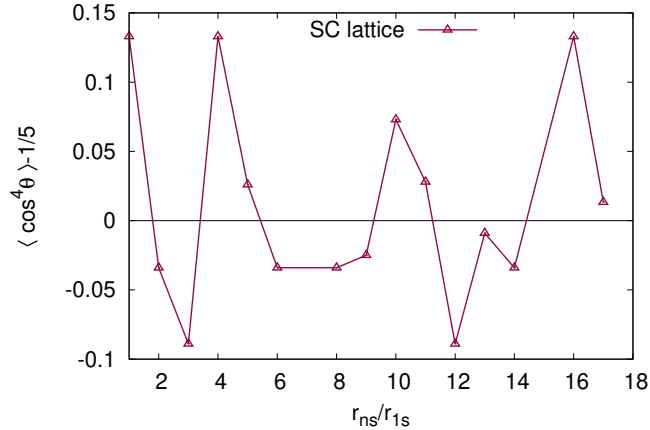}
\caption{Several first values of the average of $cos^4{\theta_{ij}}$ for the simple cubic lattice (SC) are compared with the value of the same function averaged for homogeneous particle distribution. The most significant deviations are for the coordination sphere numbers of the form $n=4^{n}$, which would happen more and more rarely as the $n$ grows. The points corresponding to the values of $n=7$ and $n=15$ are missing, as they should, due to Legendre's three-squares theorem.}
 \label{fig:oscillations}
\end{figure}

Let us consider $n$-th coordination sphere around $i$-th particle in the lattice with $N_{\mathrm{ns}}$ particles on it. The radius-vector $\bm{r}_{ij}$ which connects the particle in the center of the coordination sphere and some particle laying on it would be directed to some angle $\theta_{ij}$ with respect to the external magnetic field $\bm{H}_0$. Let us chose the magnetic field to be directed along the $Ox$ axis. Then, one may calculate the average of the $\cos^2{\theta_{ij}}$ on the $n$-th coordination sphere as follows
\begin{equation}
\langle \cos^2{\theta_{ij}} \rangle_{\mathrm{ns}} = \frac{1}{N_{\mathrm{ns}}}\sum_{j\in N_{\mathrm{ns}}} \frac{(r_{ij}^2)_x}{r^2_{ij}}.
\end{equation}
One can see that
\begin{equation}
\frac{1}{N_{\mathrm{ns}}}\sum_{j\in N_{\mathrm{ns}}} \frac{(r_{ij}^2)_x+(r_{ij}^2)_y+(r_{ij}^2)_z}{r^2_{ij}}=1.
\end{equation}
Thus, for simple cubic (SC), body centered cubic (BCC) and face-centered cubic (FCC) lattices, where all three directions are equivalent, the average of the square of cosine would be as follows
\begin{equation}
\langle \cos^2{\theta} \rangle_{\mathrm{ns}} = 1/3.
\end{equation}
This result holds for any site $i$ and any coordination sphere $n$ of the aforementioned lattices. This calculations are essentially the same as in the Note $55$ in~\cite{Lorentz1916}. The average value of the fourth power of the cosine
\begin{equation}
\langle \cos^4{\theta_{ij}} \rangle_{\mathrm{ns}} = \frac{1}{N_{\mathrm{ns}}}\sum_{j\in N_{\mathrm{ns}}} \frac{(r_{ij}^4)_x}{r^4_{ij}},
\end{equation}
does not have such a nice property for those lattices, and depends on the particular coordination sphere. In the case of SC, for example, the several first values are compared with the averaging for the homogeneous distribution $\langle \cos^4{\theta_{ij}} \rangle_{\mathrm{ns}}=1/5$ in the Fig.~(\ref{fig:oscillations}).
At first glance, the graph does not look very encouraging, but the most diverging values correspond to $n=1,4,16$ or $n=4^{n}$ and those would happen more and more rarely as $n$ grows. And if we understood the remarks found in Ref.~\cite{Bourgain2016, Duke1988} correctly, the points on unit sphere become uniformly distributed for large $n\neq0,4,7~\mathrm{mod}~8$. Average value of the $\cos^4{\theta_{ij}}$ for some lattices, up third coordination sphere can be found in the Table~\ref{tab:table5}.
\begin{table*}
	\centering
	\begin{tabular}{ |c|c|c|c|c|c|c|c|c|c| }
		\hline
		Lattice type  &   $\langle \cos^4{\theta_{ij}} \rangle_{\mathrm{1s}}$  &  $\langle \cos^4{\theta_{ij}} \rangle_{\mathrm{2s}}$    &  $\langle \cos^4{\theta_{ij}} \rangle_{\mathrm{3s}}$     &    $r_{\mathrm{2s}}$  & $r_{\mathrm{3s}}$   &   $N_{\mathrm{1s}}$   & $N_{\mathrm{2s}}$   & $N_{\mathrm{3s}}$  &      $\phi$ (cubic cell)  \\
		\hline
		SC &     $1/3$            &           $1/6$     &  $1/9$ &   $\sqrt{2}r_{\mathrm{1s}}$  &  $\sqrt{3}r_{\mathrm{1s}}$  & 6  & 12 & 8 &  $\frac{4\pi a^3}{3r^3_{\mathrm{1s}}}$\\
		BCC &     $1/9$            &           $1/3$     &  $1/6$ &   $\sqrt{4/3}r_{\mathrm{1s}}$  &  $\sqrt{8/3}r_{\mathrm{1s}}$  & 8  & 6 & 12 &  $\frac{\sqrt{3}\pi a^3}{r^3_{\mathrm{1s}}}$\\
		FCC &     $1/6$            &           $1/3$     &  $2/9$ &   $\sqrt{2}r_{\mathrm{1s}}$  &  $\sqrt{3}r_{\mathrm{1s}}$  & 12  & 6 & 24 & $\frac{4\sqrt{2}\pi a^3}{3r^3_{\mathrm{1s}}}$\\
		HCP &     $2/9$            &           $1/3$     &  $1$ &   $\sqrt{2}r_{\mathrm{1s}}$  &  $\sqrt{8/3}r_{\mathrm{1s}}$  & 12  & 6 & 2 & $\frac{4\sqrt{2}\pi a^3}{3r^3_{\mathrm{1s}}}$\\
		\hline
	\end{tabular}
	\caption{~}
	\label{tab:table5}
\end{table*}

Interestingly, for the two dimensional triangular lattice the analogous result $\langle \cos^2{\theta}\rangle=1/2$ can be obtained from the property of the particle distribution on the coordination circles of this lattice. Namely, the particles can be grouped in sets where they would be each equally spaced from their neighbors. In other words, they would be situated in the vertices of the regular polygon (in each group).
And one can see that for the polygon with $N_{\mathrm{ns}}\geq 3$ sides, the average would be as follows
\begin{equation}
\langle \cos^2{\theta} \rangle_{\mathrm{ns}}=\frac{1}{N_{\mathrm{ns}}} \sum_{k=0}^{N_{\mathrm{ns}}-1}\cos^2{(\theta_0+2\pi k/N_{\mathrm{ns}})},
\end{equation}
where $\theta_0$ is the arbitrary angle the polygon may be rotated around its center. Using power reducing formula
\begin{equation}
\langle \cos^2{\theta} \rangle_{\mathrm{ns}}=\frac{1}{2 N_{\mathrm{ns}}} \sum_{k=0}^{N_{\mathrm{ns}}-1}\big[1+\cos{(2\theta_0+4\pi k/N_{\mathrm{ns}})}\big],
\end{equation}
and rewriting the cosine as en exponent  $\cos{\theta}=\mathrm{Re}e^{i\theta}$ one may obtain the following
\begin{equation}
\langle \cos^2{\theta} \rangle_{\mathrm{ns}}=\frac{1}{2}+\frac{1}{2 N_{\mathrm{ns}}}\mathrm{Re}~e^{i 2\theta_0}\sum_{k=0}^{N_{\mathrm{ns}}-1}q^k=1/2.
\end{equation}
Here we use the geometric sum formula with $q=e^{4\pi/N_{\mathrm{ns}}}\neq 1$ and $q^{N_{\mathrm{ns}}}=1$. The similar procedure can be performed to calculate the fourth power average
\begin{equation}
\langle \cos^4{\theta} \rangle_{\mathrm{ns}}=\frac{3}{8}+\frac{1}{8N_{\mathrm{ns}}}\mathrm{Re}\sum_{k=0}^{N_{\mathrm{ns}}-1}\big[e^{ i 4\theta_0} q^{2k}+4 e^{ i 2\theta_0}q^k\big],
\end{equation}
where the $q=e^{4\pi/N_{\mathrm{ns}}}$, so the last term in parenthesis gives zero contribution for $N_{\mathrm{ns}}\geq3$. In the case of the triangular lattice, the first term in the parenthesis also vanishes, since $q^{2k}\neq 1$, as a result
\begin{equation}
\langle \cos^4{\theta} \rangle_{\mathrm{ns}}^{\mathrm{triang.latt.}}=\frac{3}{8}.
\end{equation}
If the average in each group is the same $\sum_1/N_1=\sum_2/N_2=\cdots=\alpha$, then it is also the average of the all values on the circle together $\frac{\sum_1+\sum_2+\cdots}{N_1+N_2+\cdots}=\alpha$, because $\sum_i/\sum_1=N_i/N_1=\sigma_i$ and $\frac{\sum_1}{N_1}\frac{1+\sigma_2+\cdots}{1+\sigma_2+\cdots}=\frac{\sum_1}{N_1}=\alpha$.

But more generally $q^2$ can be equal to one, for example in the square lattice, when $N_{\mathrm{ns}}=4$, and calculation of the forth power average would be coordination circle dependent, similarly as in the three dimensional case.

\end{document}